\documentclass{aa}  

\usepackage{soul}
\usepackage{xcolor}
\usepackage{graphicx}
\usepackage{multirow}
\usepackage{hhline}

\usepackage{txfonts}
\usepackage{hyperref}
\hypersetup{
    linkcolor=blue,
    citecolor=blue,
    colorlinks=true,
    filecolor=blue,      
    urlcolor=blue
    }

\usepackage{float}

\begin{document} 

   \title{Mass bias in clusters of galaxies: \\ Projection effects on the case study of Virgo replica}

   \titlerunning{Mass bias in clusters: Projection effects on the case study of Virgo replica}
   
   \author{Th\'eo Lebeau\inst{1} \thanks{E-mail: theo.lebeau@universite-paris-saclay.fr}, Jenny G. Sorce\inst{2,1,3}, Nabila Aghanim\inst{1}, Elena Hern\'andez-Mart\'inez\inst{4}, Klaus Dolag\inst{4,5}, 
          }

   \authorrunning{Lebeau et al.}

   \institute{Université Paris-Saclay, CNRS, Institut d’Astrophysique Spatiale, 91405 Orsay, France
         \and
             Université de Lille, CNRS, Centrale Lille, UMR 9189 CRIStAL, F-59000 Lille, France
        \and
            Leibniz-Institut für Astrophysik (AIP), An der Sternwarte 16, 14482 Potsdam, Germany
        \and
            Universitäts-Sternwarte, Fakultät für Physik, Ludwig-Maximilians-Universität München, Scheinerstr. 1, 81679 München, Germany
        \and
          Max-Planck-Institut für Astrophysik, Karl-Schwarzschild-Straße 1, 85741 Garching, Germany}

   \date{Received 20 July 2023 / Accepted 11 October 2023}
 
  \abstract
  {When measuring the observed pressure, density, or temperature profiles of the intracluster gas, and hence the mass of clusters of galaxies, projection effects or departures from the spherical symmetry hypothesis may induce biases. To estimate how strongly the cluster's observed properties depend on the direction of observation, we use a constrained hydrodynamical simulation of the Virgo cluster that replicates the actual cluster of galaxies. In this case study, we analysed Virgo properties when projected in different directions, including along the Milky Way-Virgo axis, which mimics our observation direction. We compared the hydrostatic mass and the hydrostatic mass bias from the projection along the different observation directions to that derived from the 3D simulation. We show that projection effects impact the determination of Virgo mass. We particularly demonstrate that the mass and pressure along the line of sight correlate with the 2D- and 3D-deprojected electron density and pressure profiles intensity and thus impact the derived hydrostatic mass. We also show that the deviations to the hydrostatic equilibrium induced by pressure discontinuities within the cluster are emphasised by the deprojection process and thus make the hydrostatic mass estimation invalid at these radii.} 

   \keywords{Galaxies: clusters: individual: Virgo - Galaxies: clusters: intracluster medium - Methods: numerical}

   \maketitle

%

\section{Introduction}

Galaxy clusters formed via the gravitational collapse of primordial over-densities and are thus the most massive gravitationally bound structures of the Universe. They are mainly composed of dark matter (DM) ($\sim 80\%$); they also contain a hot plasma trapped in the DM gravitational potential well, the intracluster medium (ICM) ($\sim 15\%$) (see \citeauthor{kravtsov2012formation} \citeyear{kravtsov2012formation} for a review). The galaxies account for about $5\%$ of the cluster's total mass. Clusters of galaxies are located at the nodes of the cosmic web \citep[e.g.,][]{peebles2020large,cautun2014evolution} and are, therefore, peaks of density in the matter distribution of the Universe. Their number count -in other words their abundance in mass and redshift- is thus a probe for the cosmological parameters (see \citeauthor{allen2011cosmological} \citeyear{allen2011cosmological} and \citeauthor{pratt2019galaxy} \citeyear{pratt2019galaxy} for reviews) provided their masses are accurately estimated. \\

We can directly infer the total cluster mass using gravitational lensing of the cluster onto background galaxies (see \citeauthor{hoekstra2013masses} \citeyear{hoekstra2013masses} for a review).
Differently, any massive cluster contains a hot ICM emitting in X-rays via bremsstrahlung \citep{sarazin1986x,bohringer2010x} that can be observed with the ROSAT \citep{bohringer2004rosat}, Chandra \citep{vikhlinin2006chandra}, XMM Newton \citep{pacaud2007xmm}, SUZAKU \citep{simionescu2014witnessing}, and eROSITA \citep{bahar2022erosita} satellites. The ICM can also be observed in the sub-millimetre domain through the inverse Compton scattering of the Cosmic Microwave Background (CMB) photons passing through the ICM (see \citeauthor{carlstrom2002cosmology} \citeyear{carlstrom2002cosmology} for a review) -the so-called thermal Sunyaev-Zel'dovich (tSZ) effect \citep{sunyaev1972observations}- using several ground-based and space telescopes: the South Pole Telescope  \citep[SPT;][]{plagge2010sunyaev}, the Atacama Cosmology Telescope \citep[ACT;][]{marriage2011atacama}, and the Planck Satellite \citep{ade2014planck}. Assuming the hydrostatic equilibrium between the ICM pressure and the gravitational potential well, we can estimate the hydrostatic mass of a galaxy cluster (see \citeauthor{ettori2013mass} \citeyear{ettori2013mass} and \citeauthor{kravtsov2012formation} \citeyear{kravtsov2012formation} for reviews). \\

The hydrostatic equilibrium hypothesis is most often invalid and leads to a so-called hydrostatic mass bias, which is defined as the ratio of the hydrostatic mass over the total mass. The value of the bias is still debated; on one hand there is a large scatter around $(1-b)=0.8$ among the values proposed by several cosmological simulations or weak lensing experiments as shown in \cite{salvati2018constraints}, \cite{gianfagna2021exploring}, and in the discussion of this work. On the other hand, the value proposed by \citep{salvati2018constraints} to reconcile the CMB and tSZ number-count constraints on the parameters $\sigma_8$, namely the amplitude of the matter power spectrum at 8~Mpc/h and the total matter density $\Omega_m,$ is $(1-b)=0.62\pm0.07$. As we enter the era of precision cosmology, we need to better understand the sources of bias in the clusters' hydrostatic mass estimation to ensure accurate measurements. \\

There are multiple possible contributions to the hydrostatic mass bias that need to be well calibrated to reach accurate mass estimates. First, we make the hypothesis of fully thermal pressure, but magnetic fields \citep[e.g.][]{dolag2000effect}, cosmic rays \citep[e.g.][]{boss2023crescendo}, or turbulence \citep[e.g.][]{rasia2004dynamical,nelson2014hydrodynamic,pearce2020hydrostatic} in the ICM could add significant non-thermal pressure. Second, we assume that clusters are spherical and relaxed although asphericity and unrelaxedness of clusters \citep[e.g.][]{gouin2021shape} could also contribute to the bias. Moreover, temperature structure in the ICM \citep[e.g.][]{rasia2006systematics,rasia2012lensing}, the introduction of an inertia term in the mass estimation \citep[e.g.][]{suto2013validity,biffi2016nature}, redshift evolution of the bias \citep[e.g.][]{wicker2023constraining}, or the cluster's local environment impact on the hydrostatic equilibrium \citep[e.g.][]{gouin2022gas,gouin2023soft,vurm2023cosmic} might also need to be accounted for. In addition to these physical sources of biases, using projected quantities (accessible via observations) instead of 3D ones (accessible in numerical simulation) introduces other biases. Furthermore, since we can only observe the ICM of clusters from our unique position in the Universe, reconstructed thermodynamical properties of clusters can then be affected by the presence of gas along our line of sight (LoS).\\ 

For the Virgo cluster replica case study, our work focuses on two main aspects. We show how the physics in the cluster core and the environment in the cluster outskirt impact the hydrostatic mass and the bias. We also explore how the signatures of these physical conditions, and hence the mass bias, are affected by the projection in different directions. The impact of projection effects on cluster mass estimates was studied by, for instance, \citet{ameglio2007joint,ameglio2009reconstructing},\citet{meneghetti2010weighing}, and \citet{barnes2021characterizing} based on mock observations of the tSZ, X-rays, or lensing signal from low-resolution simulations. In our study, we used direct simulation 3D and projected quantities to identify the main sources of biases. The analysis of mock data and the comparison with actual observations will be presented in a forthcoming paper. Additionally, we used a constrained cosmological simulation for the first time to quantify the projection effects by observing a cluster replica from multiple directions, including our observer LoS. Our high-resolution, hydrodynamical-simulation -including supernova (SN) and active galactic nucleus (AGN) feedback- replica of the Virgo cluster \citep{sorce2021hydrodynamical} allowed us to explore the impact of the feedback in the cluster core and its impact on the mass bias.\\

This paper is organised as follows: we describe the Virgo cluster and its simulated replica in Section \hyperref[sec:2]{2}. The 3D, 2D, and 3D-deprojected radial profiles are presented in Section \hyperref[sec:3]{3}, and the derived hydrostatic mass bias is discussed in Section \hyperref[sec:4]{4}. We discuss the projection effects in Section \hyperref[sec:5]{5}. We discuss the impact of fitting the radial profiles and the addition of non-thermal pressure on the hydrostatic mass estimation and compare our mass biases to the literature in Section \hyperref[sec:6]{6}. We draw our conclusions in Section \hyperref[sec:7]{7}.

\section{The Virgo cluster}
\label{sec:2}

\subsection{Observed properties of the cluster}

The Virgo cluster is our closest neighbour, located more or less 16~Mpc away from us \citep{mei2007acs} at a redshift of 0.0038 \citep{wu1998updating}. Table \ref{tab:mass} shows its mass obtained from different observations. Our proximity to this galaxy cluster permitted us to study its galaxy populations \citep{de1961structure} and their dynamics within this very dense environment in detail (e.g. \cite{karachentsev2014infall} using the Hubble Space Telescope (HST), the GALEX Ultraviolet Virgo Cluster Survey
(GUViCS) from \cite{boselli2016galex}, the Herschel Virgo cluster surveys of \cite{pappalardo2015herschel}, the Next
Generation Virgo Cluster Survey (NGVS) of \cite{ferrarese2016next}, or the Virgo Environmental Survey Tracing Ionised Gas Emission (VESTIGE) of \cite{boselli2018virgo}. The filaments connected to the Virgo cluster have also been studied in detail \citep{kim2016large,lee2021properties,castignani2022virgo,castignani2022virgoII}. The Virgo ICM has also been inspected using its tSZ signal \citep{planck2016virgo} and its X-ray emission from the core \citep{young2002chandra,simionescu2017witnessing} to the virial radius \citep{urban2011x,simionescu2015uniform}, providing information about the pressure, electron density, temperature, and metallicity distributions. In addition, the Virgo cluster recently received a lot of attention due to the observation of the black hole (BH) of its central galaxy M87 by the Event Horizon Telescope Collaboration \citep{collaboration2019first}. This BH is an AGN expelling gas in a jet mode \citep{lucchini2019unique,ghizzardi2004radiative}. \\

Accurately determining the complex formation history of Virgo is challenging. However, \cite{boselli2008origin} estimated that in the past few gigayears, Virgo accreted around 300 galaxies with $M_{\star}>10^7M_{\odot}$ per gigayear. Moreover, \cite{lisker2018active} found that a group of about $10\%$ of the cluster mass would have recently fallen into the cluster. This agrees with the large number of substructures observed within the cluster \citep{huchra1985virgo,binggeli1987studies}, showing that Virgo is an unrelaxed and dynamically young cluster.

\subsection{Simulated replica}

Numerical simulations are used to improve our understanding of past and current Virgo properties. Hence, several projects \citep{hoffman1980dynamical,li2014modeling,moran2014globular,zhu2014next} have proposed numerical models of the cluster as an isolated object, thus missing the impact of Virgo's environment on its build up and its properties. 

More recently, \cite{sorce2019virgo} ran 200 DM simulations of the Virgo cluster. For the very first time, these simulations reproduce the cluster local environment, including filaments connected to Virgo and groups of galaxies in its vicinity, using initial conditions of the local Universe constrained with both observed galaxy positions and peculiar velocities but no mass-luminosity relation \citep{sorce2016cosmicflows}. The masses of the cluster replicas in the simulations thus result from the evolution of the initial fields without any cluster mass calibration. The dynamical state and formation history of these simulated replicas are in excellent agreement with the observed properties of the Virgo cluster \citep{olchanski2018merger}. They show a dynamically young and unrelaxed object with a mean offset of the centre of mass with respect to the spherical centre of $58\pm38$~kpc and $79\pm22$ substructures with a mass above $10^{10}$M$_{\odot}$ (see \citeauthor{sorce2019virgo} \citeyear{sorce2019virgo} for more details). \\

Among the 200 DM simulations, the most representative halo (i.e. the closest to the average properties of the full sample, with a merging history in agreement with the mean history of the total sample) was re-simulated, including baryons using the adaptive mesh refinement code RAMSES \citep{teyssier2002cosmological}. Our paper uses this unique, high-resolution, zoom-in hydrodynamical simulation of Virgo, the Constrained LOcal and Nesting Environment (CLONE) simulation, whose main properties are briefly described hereafter (for more details, see \citeauthor{sorce2021hydrodynamical} \citeyear{sorce2021hydrodynamical} and references therein). For this run, the Planck cosmology from \cite{planckcosmoparam2014} was adopted with spectral index $n_s=0.961$, total matter density $\Omega_m=0.307$, dark energy density $\Omega_{\Lambda}=0.693$, baryonic density $\Omega_b=0.048$, Hubble constant $H_0=67.77~\mathrm{km/s/Mpc}$, and amplitude of the matter power spectrum at 8~Mpc/h $\sigma_8 = 0.829$. The zoom region, contained in a 500~Mpc/h local Universe box \citep{sorce2016cosmicflows}, is a 30~Mpc diameter sphere with a resolution of $8192^3$ effective DM particles of mass $m_{\mathrm{DM}}=3.10^7~\mathrm{M_\odot}$. On the baryon side, the finest cell size is 0.35~kpc. In this work, we also used a low-resolution simulation containing $2048^3$ effective DM particles with $m_{\mathrm{DM}}=2.10^9~\mathrm{M_\odot}$ and a finest cell size of 1.4~kpc. The sub-grid physical models for radiative gas cooling and heating, star formation, kinetic feedback from type II SNs and AGNs follow the Horizon-AGN implementation of \cite{dubois2014dancing,dubois2016horizon}. Moreover, the AGN feedback model has been improved by orientating the jet according to the spin of the BH (see \citeauthor{dubois2021introducing} \citeyear{dubois2021introducing} for details). \\

Overall, the simulation is in good agreement with observations on several aspects. First, the halo population of the cluster is consistent with observations, the luminosity function derived from the simulation matches well with that obtained from observations up to the completeness threshold of the simulation, that is $10^9~\mathrm{M_{\odot}}$ for the low-resolution and $10^{8.5}~\mathrm{M_{\odot}}$ for the high-resolution simulation\ \citep[see Appendix D in][]{sorce2021hydrodynamical}. Then, the dynamical galaxy distribution matches with observations, giving consistent virial and zero velocity radii\footnote{The radius at which the mean radial velocity of galaxies is zero; see Appendix C of \cite{sorce2021hydrodynamical} for more details.}. Moreover, the merging history, the central galaxy M87-replica (albeit a bit more massive than the actual M87) or the filamentary structure are also in good agreement with observations (see \citeauthor{sorce2021hydrodynamical} \citeyear{sorce2021hydrodynamical} for more details). Given the good agreement between the Virgo replica and the actual cluster, we are confident that the former can be used to study observational and physical effects impacting the mass estimate. \\ 

We used the rdramses\footnote{\url{https://github.com/florentrenaud/rdramses}.} code to extract the position, velocity, mass, level of resolution, electron density, pressure, and temperature of each cell from the simulation output files. We also used this code to extract the DM particles' positions, velocities, and masses. To identify the Virgo cluster in the large-scale simulation, DM halos have been identified using the halo finder described in \cite{tweed2009building}, where galactic halos are identified using the stars as tracers  \citep{tweed2009building}. The Virgo DM halo has a characteristic radius\footnote{The radius within which the mean matter density is 500 times the critical density of the Universe, defined as $\rho_c=\frac{3H^2}{8\pi G}$. The actual critical density is $\rho_{c,0}=\frac{3H_0^2}{8\pi G}$.} $R_{500}=1087~\mathrm{kpc}$, the total mass contained in the sphere of this specific radius is $M_{500}=3.36~10^{14}~\mathrm{M_{\odot}}$. Its virial radius is $R_{vir}=2147~\mathrm{kpc}$ and the mass in this radius is $M_{vir}=6.31~10^{14}~\mathrm{M_{\odot}}$. Our study focuses on the ICM gas, and we thus selected the cells with a temperature above $10^5~\mathrm{K}$ and removed the cells associated with the galaxies. More precisely, we removed the cells within the virial radius of the DM halo associated with the galaxies with a mass above the completeness threshold of the simulation. We only keep the cluster's central galaxy, M87, its DM halo being the cluster halo.\\

\subsection{Pressure and electron density maps}
\label{subsec:2.3}

In this work, rather than using mock observations of tSZ or X-rays, we studied the cluster using direct simulation 3D outputs of the pressure and electron density that we compared to 3D-deprojected quantities obtained following different directions. Thanks to the accuracy of the Virgo replica together with its local environment up to $\approx$ $7*R_{vir}$, and to the implementation of SN and AGN feedback, we were able to carry out an unprecedented study of projection effects on the hydrostatic mass estimation of the Virgo cluster; that is to say, we investigated the combined impact on the hydrostatic mass of the signature of feedback processes (i.e. shocks) and the matter distribution along the LoS. Therefore, our approach does not exclude any substructure along the projection directions. Additionally, using the Virgo constrained simulation allows us to compare the projected quantities with the actual observing LoS directly. \\

\begin{figure}
   \centering
    \includegraphics[width=\hsize]{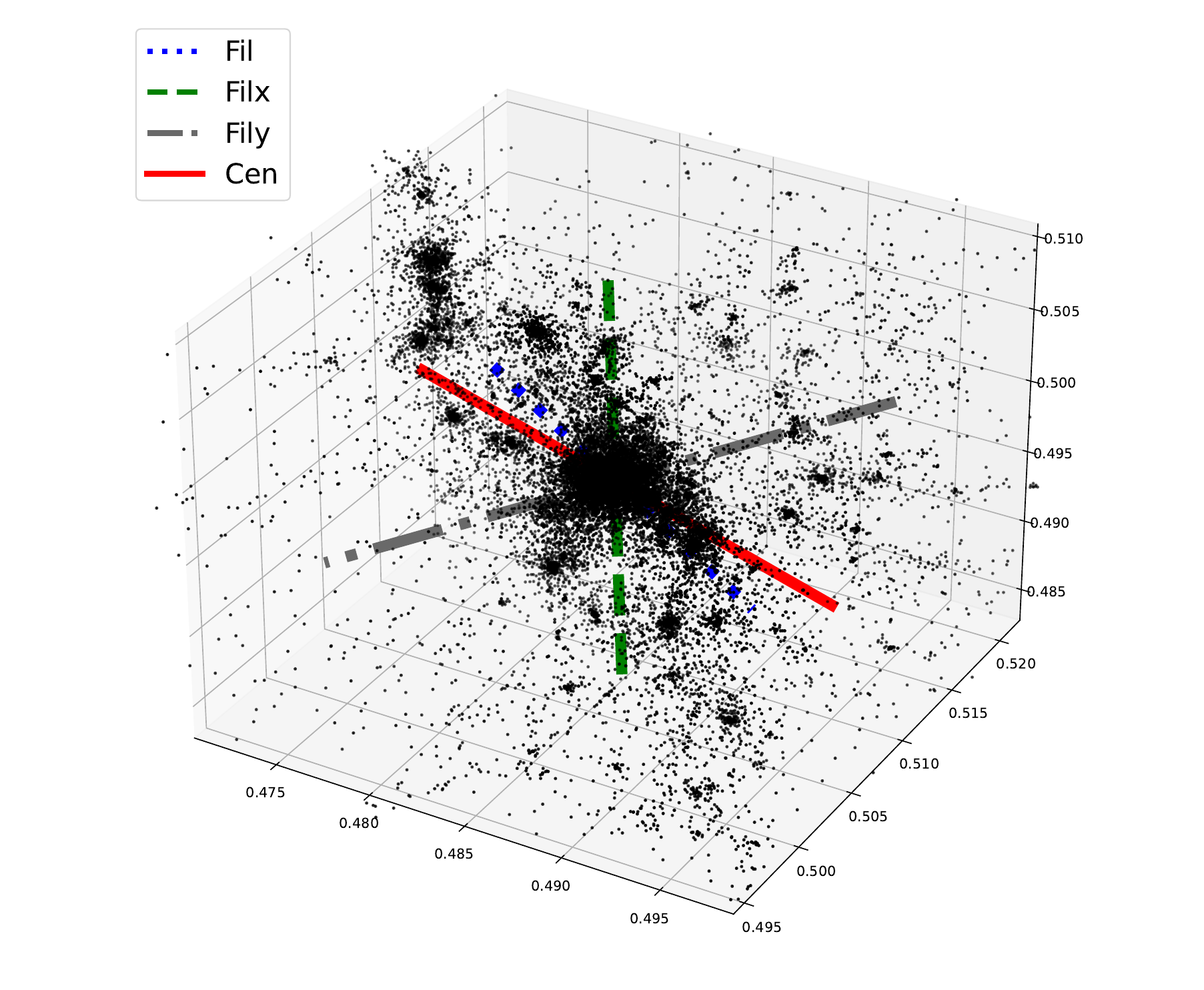}
      \caption{\label{3D_box} DM box of simulated Virgo cluster. The coloured lines represent the axis of projection. The red solid line axis is for the projection from the centre of the box to Virgo (Cen), the blue dotted line shows the projection along the main filament (Fil), and the green dashed and grey dash-dotted lines show the projections perpendicular to the main filament, with a rotation around the x (Filx) and y (Fily) axis, respectively}     
   \end{figure}

Four directions are considered in our study (shown in Fig. \ref{3D_box}) that are associated with projected maps. The projection along the red LoS is from the centre of the large-scale simulation box to the Virgo cluster (Cen hereafter, solid red). The large-scale simulation is almost centred on the Milky Way \citep{sorce2016cosmicflows}; we thus assume that this projection is comparable to the real observation of the Virgo cluster, as shown in \cite{sorce2023velocity} in their study of velocity waves in the Hubble diagram. Since we studied the impact on the hydrostatic mass estimation of the gas distribution along the LoS, we projected along the main filament connected to Virgo, assuming it would contain a lot of mass along the LoS (the blue dotted LoS in Fig. \ref{3D_box}; Fil hereafter). We also projected the quantities orthogonally to the main filament; these are two extreme cases with less mass along the LoS represented as the green and grey LoS in Fig. \ref{3D_box}. Both are perpendicular to the main filament and, respectively, follow a rotation around the x- (Filx, in green dashed) and y-axes (Fily; dash-dotted grey).  \\

The maps of pressure and electron density are produced by summing the contributions of the cells along the LoS in each pixel. The cells contribute to one or more pixels depending on their resolution level. For the low-resolution zoom-in simulation ($2048^3$ effective DM particles), the pixel size equals the finest cell size (i.e. 1.4~kpc). For the high-resolution zoom-in simulation ($8192^3$ effective DM particles), among the cells in highly refined areas (i.e. within the central 100~kpc), only some cells reach the finest resolution level (less than $15\%$), and about $60\%$ of them have a cell size of 1.4~kpc at best. Outside the cluster core, some cells reach the highest resolution level in the galaxies, but they have been removed during the ICM cell selection described above. So, we fixed the resolution level of the cells -and so the size of the pixels- to 1.4~kpc when running rdramses. We performed tests with maps with pixel sizes equal to the finest cell size (i.e. 0.35~kpc) and found very similar projected quantities. The mean relative difference on 2D radial profiles was $6.4 \times  10^{-5}$. \\

\begin{figure*}
   \centering
    \includegraphics[trim=250 50 550 100,width=1\textwidth]{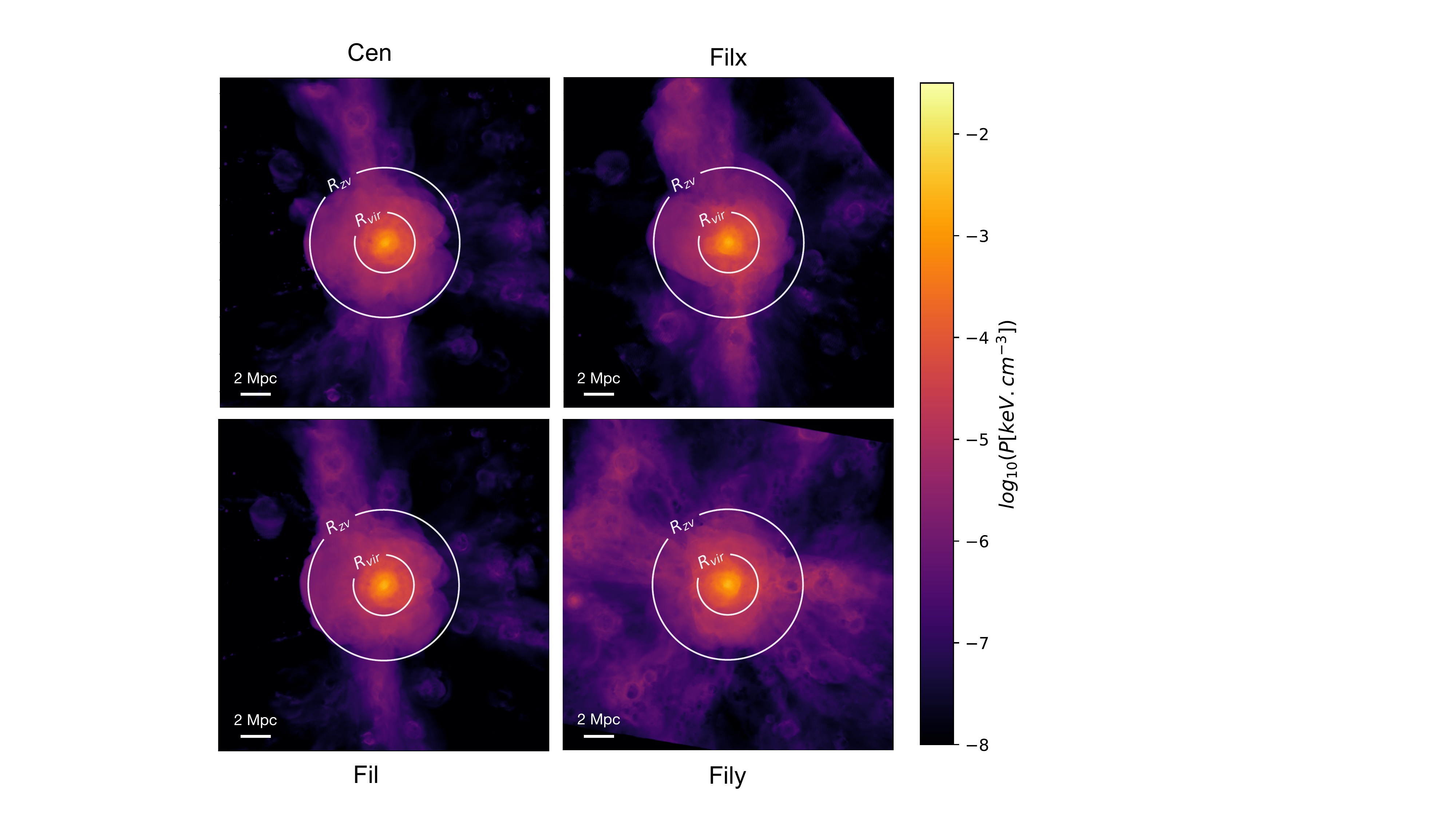}
      \caption{\label{p maps 4} Pressure maps in four directions from the high-resolution zoom-in simulation. The maps are 22.123~Mpc wide and contain $15728^2$ pixels. The central circle is the virial radius, the outer one is the zero velocity radius. The colour scale ranges from $10^{-8}$ to $10^{-1.5}$ kev.c$\mathrm{m^{-3}}$. From top to bottom and left to right, the Cen, Fil, Filx, and Fily projections associated with directions shown in Fig. \ref{3D_box}. These maps show the matter distribution in Virgo's local environment. The cluster is part of a matter sheet (visible on the Fily projection) including a main filament (top of the Filx projection) and three secondary filaments (bottom of the Filx projection and top and bottom of the Fil and Cen projections).}
         
   \end{figure*}

The pressure maps are built using the mass-weighted mean of the cells in each pixel because the pressure is an intensive quantity. The electron density maps are column density maps, that is the sum of the electron density of the cells multiplied by their size because it is an extensive quantity. We present the pressure maps along the four directions in Fig. \ref{p maps 4}. These square maps are 22.123~Mpc wide, centred on the cluster's centre of mass, and contain $15728^2$ pixels. We show characteristic radii of the cluster; the central circle is the virial radius ($R_{vir}$), the outer one is the zero velocity radius \citep[$R_{zv}$, from][]{sorce2021hydrodynamical}. \\

These maps show several features, both at large and small scales. At large scales, we observe the matter distribution in the Virgo environment. The Fily map shows a large amount of high pressure (above $10^{-5}$ $\mathrm{keV.cm^{-3}}$) beyond the zero velocity radius. In contrast, the high pressure is concentrated in filamentary structures for the three other maps. The Fil map shows a thinner vertical filament than the main one visible on the Filx map. It shows that the matter is concentrated in a sheet crossing the cluster, which is also visible on the Fily map. Moreover, the maps show differences in the cluster shape within the zero velocity radius; e.g. the Filx-map shows an elongated cluster, and the spherical symmetry hypothesis is therefore strongly challenged in this case.\\

 Within the virial radius of the Filx map, there is a pressure discontinuity perpendicular to the main filament connected to the cluster. This is due to the accretion of the warm gas falling from the filament. This gas is heated while entering the cluster (see \citeauthor{gouin2022gas} \citeyear{gouin2022gas}, \citeauthor{gouin2023soft} \citeyear{gouin2023soft} and \citeauthor{vurm2023cosmic} \citeyear{vurm2023cosmic} and references therein for recent studies) and is then shocked (see \citeauthor{markevitch2007shocks} \citeyear{markevitch2007shocks} for a review on shocks in clusters). In the cluster's centre, we observe another pressure discontinuity, which is more or less elliptical depending on the projection. This is due to the AGN feedback of the cluster's central galaxy, M87. In this simulation, the AGN jet direction depends on the galaxy spin that is mainly perpendicular to the Filx direction. The AGN feedback is thus oriented mainly in the Filx direction, so the pressure discontinuity is seen spherical in this projection (as we observe on the top right panel of Fig. \ref{p maps 4}). Consequently, in the other projections, we observe an elliptical pressure discontinuity. This pressure discontinuity has been observed in the X-rays; the associated jet power has been estimated to be $F_{jet}\approx 3\times 10^{42} \mathrm{erg.s^{-1}}$ \citep{young2002chandra}. \\

\section{Radial profiles}
\label{sec:3}

In this section, we first show the 3D radial profiles computed from the simulation box. We then derive 2D radial profiles extracted from the maps and finally estimate 3D-deprojected profiles that we compare to the 3D ones. 

\subsection{3D radial profiles} 
\label{sub sec:3.1}

\begin{figure}
    \centering
    \includegraphics[width=0.5\textwidth]{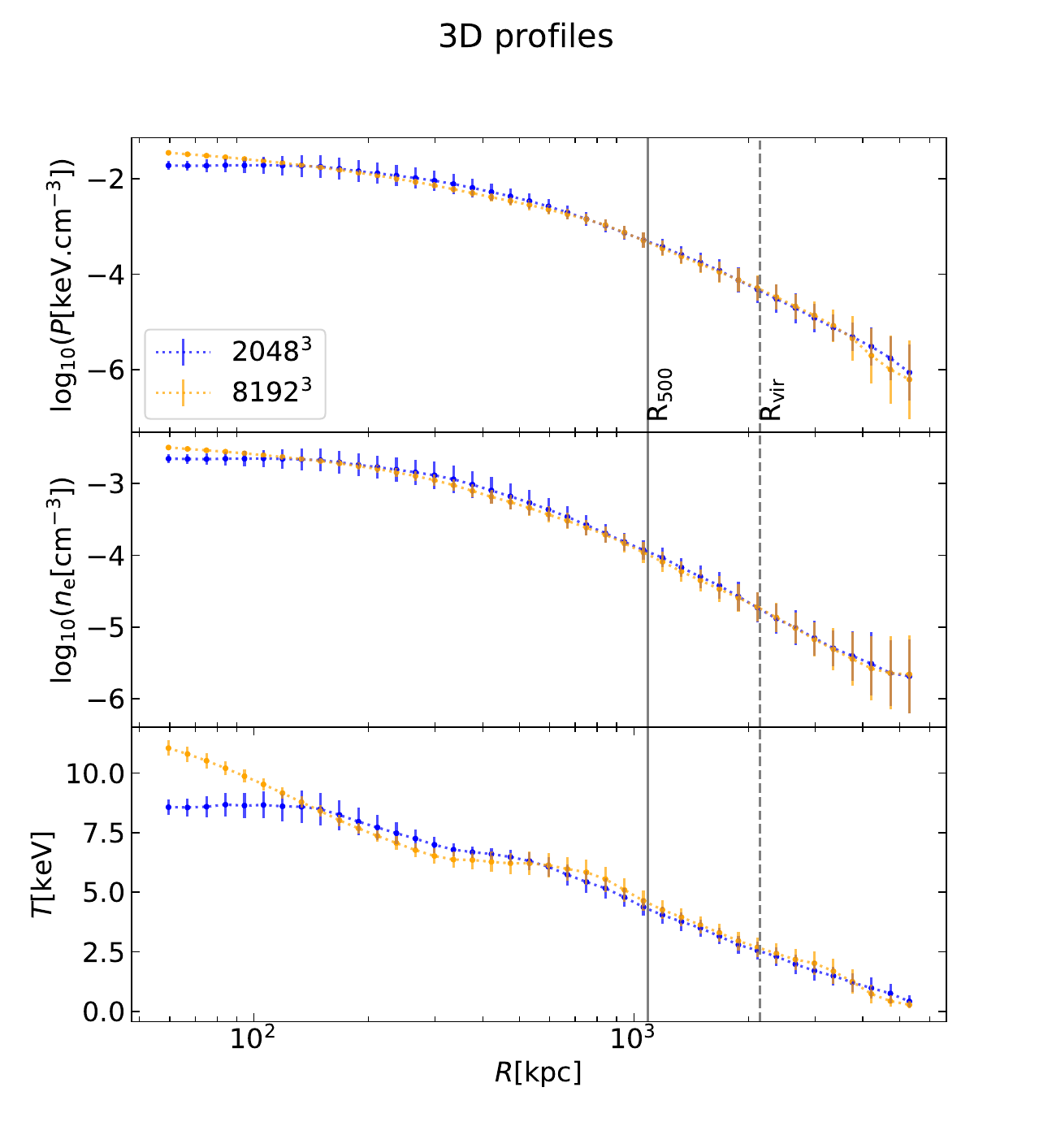}
    
    \caption{\label{3D_profs} 3D radial profiles of mass-weighted mean of pressure (top, in $\mathrm{keV/cm^3}$), electron density (middle, in $\mathrm{cm^{-3}}$), and temperature (bottom, in keV). The uncertainties are the mass-weighted standard deviation. The orange dotted lines stand for the high-resolution simulation ($8192^3$), and the blue dotted lines stand for the low-resolution simulation ($2048^3$). The profiles extend from 60 to 5300~kpc. The vertical solid line is for $R_{500}$, and the dashed vertical line is for $R_{vir}$. Profiles derived from the high- and low-resolution simulations are consistent, except in the cluster core. The low-resolution simulation has converged outward from at least R$_{500}$.}
    
\end{figure}

We computed the 3D radial profiles of the pressure, electron density, and temperature using the simulation cells. We computed the mass-weighted mean of each quantity in 40 bins of 0.05 length in logarithmic scale, ranging from 56~kpc to 5.623~Mpc. The uncertainties are given by the mass-weighted standard deviation. They are shown in Fig. \ref{3D_profs}. In each figure, the orange dotted lines stand for the high-resolution simulation ($8192^3$) and the blue dotted lines stand for the low-resolution simulation ($2048^3$). The solid grey vertical line represents $R_{500}$, the dashed one is the virial radius $R_{vir}$, and the horizontal axis is the radius in kpc ranging from 60 to 5300~kpc. \\

\subsubsection{Pressure profiles}

First, the pressure profiles are shown at the top panel of Fig. \ref{3D_profs}. In the inner cluster region within the first 150~kpc, the pressure from the high-resolution simulation is larger than the low-resolution one; the mean relative difference is $\sim 10\%$. It is reasonable, given that the low-resolution simulation does not resolve the small-scale physics in the cluster's centre, the low-resolution simulation has not converged in this area. Beyond 150~kpc, the two profiles are very similar, the mean relative difference between high and low-resolution in this range is $\sim 2.0\%,$ and both profiles agree within the error bars, particularly in the range [900,2500]~kpc. This shows that the low-resolution simulation reaches convergence beyond 150~kpc. \\

In Section \ref{sec:6}, we discuss the difference between the hydrostatic mass obtained from simulation-derived and fitted profiles. To this aim, we fit the pressure profile to the generalised Navarro-Frenk-White (gNFW) model \citep{nagai2007effects} following the notation from the \cite{ade2013planck}. The radial profile is normalised at $R_{500}$ as well as the pressure, we have

\begin{equation}
  \mathbb{P}(x)=\frac{P(r)}{P_{500}}  \quad \text{with} \quad x=\frac{r}{R_{500}}
.\end{equation}

The gNFW model is then written as

\begin{equation}
\mathbb{P}(x)=\frac{P_0}{\left(c_{500} x\right)^\gamma\left[1+\left(c_{500} x\right)^\alpha\right]^{(\beta-\gamma) / \alpha}}.
\end{equation}

This model is defined by five parameters: the normalisation, $P_0$, the concentration at $R_{500}$, $c_{500}$, and three parameters for the slope of the profile, which are  $\gamma$ for the core, $\alpha$ for the intermediate radii, and $\beta$ for the outskirts. The best-fit parameters are given in Table \ref{tab:p_fit} for 3D and 3D-deprojected profiles. The priors are the values from \cite{ade2013planck} ($P_0$=6.41, $c_{500}$=1.81, $\alpha$=1.33, $\beta$=4.13, $\gamma$=0.31) and the bounds are, respectively, [0,600], [0,20], [0,10], [0,50], and [-10,10]. \\

\begin{table*}[h]
    \centering
    \caption{Best-fit parameters of the gNFW model for the 3D high- and low-resolution simulation and 3D-deprojected pressure radial profiles from the high-resolution simulation.}
    \renewcommand{\arraystretch}{1.2}
    \begin{tabular}{ c c c c c c }
    \hline \hline
        Profile & $P_0$ & $c_{500}$ & $\alpha$ & $\beta$ & $\gamma$ \\
        \hline
        3D $8092^3$ & 600$\pm$3781 & 1.82e-02$\pm$0.11 & 0.52$\pm$0.25 & 27.3$\pm$59.6 & -0.12$\pm$0.59 \\
        3D $2048^3$ & 600$\pm$451 & 2.16$\pm$0.11 & 0.95$\pm$0.09 & 5.09$\pm$0.20 & -0.95$\pm$0.24 \\
        Fil & 22.9$\pm$20.7 & 2.00$\pm$0.33 & 2.09$\pm$0.95 & 4.32$\pm$0.32 & 0.59$\pm$0.43\\
        Filx & 600$\pm$5812 & 0.47$\pm$0.96 & 0.69$\pm$0.63 & 9.22$\pm$8.04 & -0.32$\pm$1.71 \\
        Fily & 600$\pm$6781 & 0.33$\pm$1.01 & 0.67$\pm$0.70 & 10.5$\pm$12.5 & -0.26$\pm$1.85 \\
        Cen & 13.3$\pm$15.38 & 1.83$\pm$0.54 & 3.88$\pm$5.55 & 4.24$\pm$0.62 & 0.16$\pm$0.68 \\
    \hline
    
    \end{tabular}
    \label{tab:p_fit}
\end{table*}

\subsubsection{Electron density profiles}

In the middle panel, we show the electron density profiles. We observe the same trend as for the pressure profiles for $R<150$~kpc. The mean relative difference in this region is about $3.6\%$. Then, in the [200,900]~kpc range, the high-resolution profile decreases more sharply than its low-resolution counterpart, but it is still within the error bars (the mean relative difference beyond 150~kpc is $\sim 1.3\%$). Beyond 900~kpc, the two resolutions give consistent results. We notice a flattening of the profiles at very large radii beyond 4~Mpc. At this distance, we reach the boundaries of the cluster; this plateau is due to the matter around the cluster in the cosmic filaments connected to it, and they are visible on the projected maps (see Fig. \ref{p maps 4}). \\

For the electron density, we fitted the radial profile to the isothermal $\beta$ model first introduced by \cite{cavaliere1976x,cavaliere1978distribution} and written as 

\begin{equation}
n_{\mathrm{e}}(r)=n_{\mathrm{e}}(0)\left[1+\left(\frac{r}{r_{\mathrm{c}}}\right)^2\right]^{-3 \beta / 2}
,\end{equation}where $n_e(0)$ is the central electron density, $r_c$ is the core radius, and $\beta$ is the parameter of the model. The best-fit parameters are given in Table \ref{tab:ne_fit}. \\

\begin{table}[h]
    \centering
    \caption{Best-fit parameters of the $\beta$ model for the electron density radial profiles for high- and low-resolution simulation 3D profiles and 3D-deprojected profiles from high-resolution simulations.}
    \renewcommand{\arraystretch}{1.2}
    \begin{tabular}{c c c c}
    \hline \hline
        Profile & $n_e(0)$ ($\times 10^{-3}$) & $r_c$ ($\times 10^{2}$) & $\beta$ ($\times 10^{-1}$)\\
        \hline
        3D $8092^3$ & 2.88$\pm$0.09 & 3.01$\pm$0.09 & 8.62$\pm$0.11 \\
        3D $2048^3$ & 2.46$\pm$0.05 & 3.75$\pm$0.08 & 9.21$\pm$0.08 \\
        Fil & 3.24$\pm$0.17 & 2.69$\pm$0.13 & 8.40$\pm$0.14 \\
        Filx & 2.78$\pm$0.17 & 2.87$\pm$0.16 & 8.59$\pm$0.17 \\
        Fily & 2.72$\pm$0.17  & 2.97$\pm$0.17 & 8.87$\pm$0.19 \\
        Cen & 3.13$\pm$0.26 & 3.08$\pm$0.24 & 9.01$\pm$0.26 \\
    \hline
    
    \end{tabular}
    \label{tab:ne_fit}
\end{table}

\subsubsection{Temperature profiles}

We show the temperature profiles in the bottom panel of Fig. \ref{3D_profs}. Similarly to the previous profiles, the two resolutions are consistent at large radii from 400~kpc to 4~Mpc (mean relative difference of $5.6\%$). However, in the centre of the cluster, the difference is large mainly because, contrary to the profiles in the upper panels, the profile is shown in linear scale, and the mean relative difference in the inner 80~kpc is $20.5\%$. We observe that the low-resolution profile flattens, whereas the high-resolution one becomes steeper and reaches more than 10 keV in the most central bin. The temperature being computed as the ratio of the pressure over the electron density from the simulation, this result is coherent with our conclusions on the pressure and electron density profiles.

\subsection{2D profiles}
\label{sub-sec:3.2}

We derived 2D radial profiles from the maps described in Sect. \ref{sec:2}. For that, we defined circular annuli centred on the cluster's centre, which is also the map's centre. We calculated the mean value of the given quantity in each annulus. The deprojection method is run 100 times (as described in the following subsection), for this figure the 2D radial profiles are interpolated and extrapolated so that they have the same binning. We compute the mean of the 100 realisations, the error bars are the dispersion of the sample. \\

\begin{figure*}
        \begin{minipage}[s]{1\textwidth}

            \centering
            \includegraphics[width=.49\textwidth]{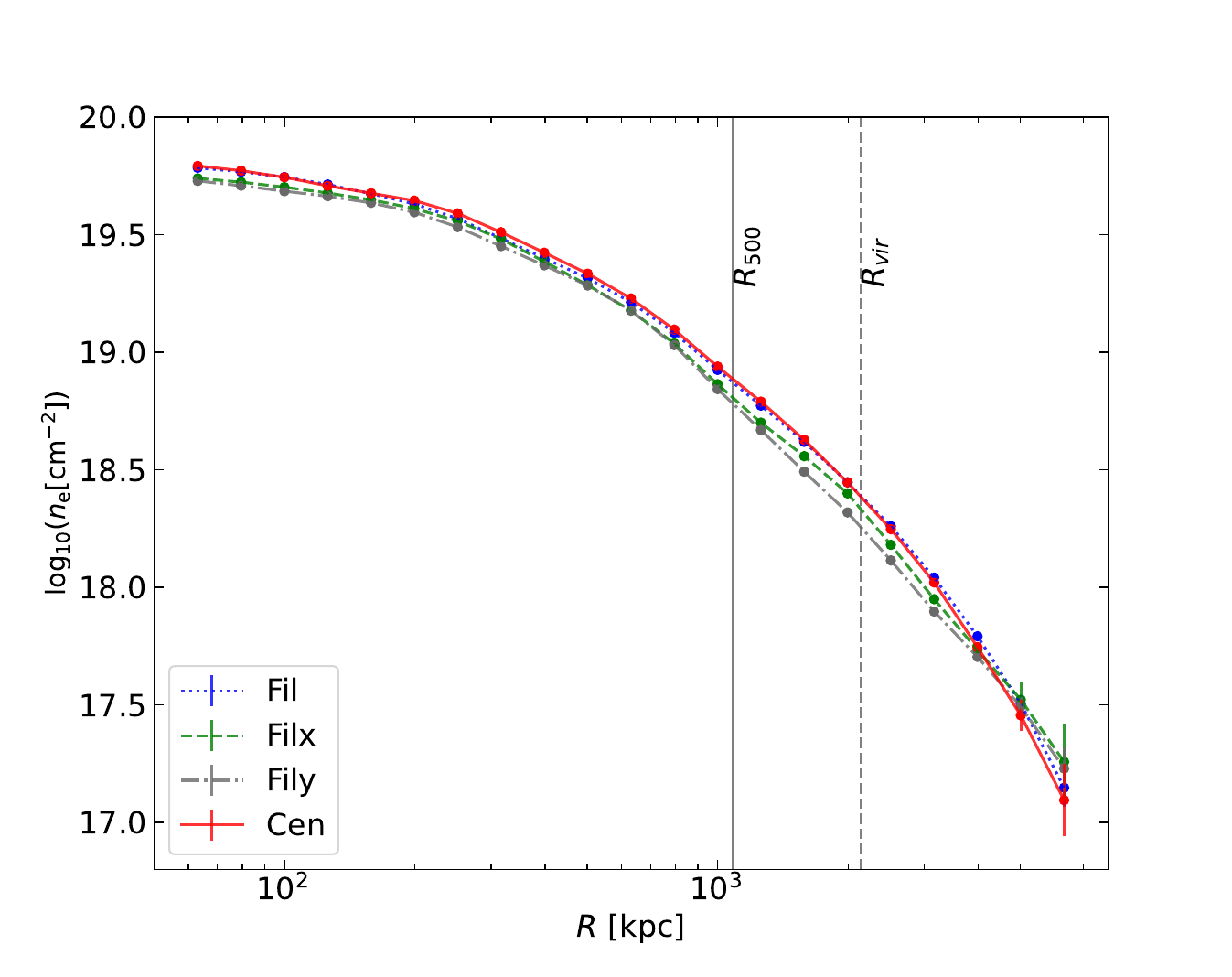}
            \hspace{0.1cm}
            \includegraphics[width=.49\textwidth]{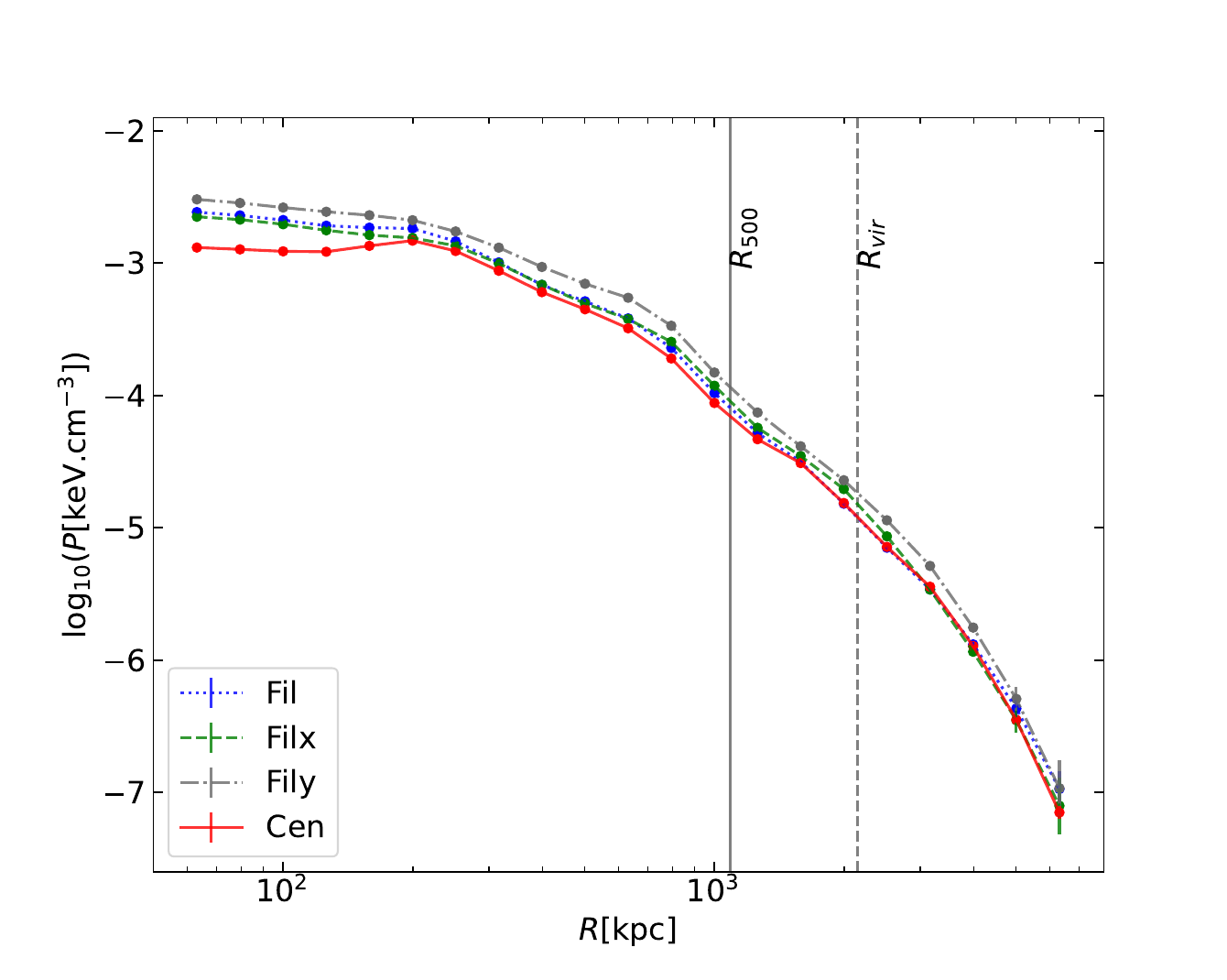}
            \caption{2D radial electron density (left) and pressure (right) profiles. The blue dotted profile is the Fil projection, the green dashed and grey dash-dotted profiles are the Filx and Fily projections, respectively. The solid red profile is the Cen projection. The vertical solid line is $R_{500}$, the vertical dashed line is the virial radius. The electron density profiles intensity is proportional to the total matter in each direction. More mass along a given direction outside the cluster induces a lower mean pressure.}
            \label{proj_profs}
       \end{minipage}
\end{figure*}

In Fig. \ref{proj_profs}, we present the 2D radial profiles of electron density (left) and pressure (right). Similarly to Fig. \ref{3D_box}, the blue dotted profile is the Fil projection, the green dashed and grey dash-dotted profiles are the Filx and Fily projections. The solid red profile is the Cen projection. The vertical solid line is $R_{500}$, the vertical dashed line is the virial radius. At first order, the electron density profiles show the same trend from the outskirt to $R_{500}$. We observe the same for pressure profiles except for the central part of the cluster where deviations arise. We discuss the projection effects on the 2D radial profiles in Sect. \ref{sec:5}.

\subsection{3D-deprojected profiles} \label{deproj_sub}

In observations, we can only access projected cluster quantities and use deprojection techniques to retrieve 3D quantities. In this work, we adapted the non-parametric geometrical iterative deprojection method proposed by \cite{mclaughlin1999efficiency}. It was initially developed for deprojecting stellar globular cluster density, but it has been commonly used for the study of the ICM in galaxy clusters \citep[e.g.][]{ameglio2007joint,ameglio2009reconstructing,tchernin2016xmm,shitanishi2018thermodynamic,ghizzardi2021iron}.\\

The circular annuli on the map are cylinders along the LoS. Assuming that the cluster is perfectly spherical, we can decompose it in spherical shells with the same binning (see Fig. \ref{dep_scheme} in Appendix \ref{appendix_algo} for a visualisation). Therefore, the signal in the cylinders is the sum of the contributions of the spherical shells along the LoS. Assuming that the background is also properly removed, the signal in the last cylinder is equal to that in the last spherical shell. In the second-to-last cylindrical shell, we have the contribution of the last spherical shell and the second-to-last one, knowing the signal in the former we can retrieve that in the latter by subtraction. With this approach, we can iteratively derive the signal in each spherical shell from the outskirts to the centre of the cluster. For the electron density, we used the method as proposed by \cite{mclaughlin1999efficiency}, except for the background. For the pressure, given that it is an intensive quantity contrary to the extensive electron density, we modified the method. The details of the deprojection technique can be found in Appendix \ref{appendix_algo}. \\

For the present study, we also adapted the background subtraction for each quantity. For the ICM pressure, the background signal is evaluated as the mean value of the pixels within an interval ranging from 8 to 10~Mpc from the centre of the map. This interval is chosen to remove the local environment contribution without removing signal from the cluster. The background is evaluated far enough from the cluster, between $\approx$ 4 and 5 $R_{vir}$, so that it does not include the filaments connected to the cluster and is subtracted from the 2D radial profile bins. For the electron density, the background is evaluated as the mean electron density in the [8,10]~Mpc range divided by the depth of the line of sight (22.123~Mpc in our case). The relative volume of the sphere in the cylinders varies at each bin, and so does the volume of the background, we thus subtracted it at each step of the iterative process (see Appendix \ref{appendix_algo} for scheme and details). \\

The deprojection method was run 100 times while varying the lower and upper radial bounds of the 2D radial profile. The lower bound is drawn from a uniform distribution in log scale in the [50,100]~kpc interval; we did the same for the upper bound in the [4,6]~Mpc interval. In the cluster's core, the size of the cells is approximately 0.1 in log scale, so, to avoid under-sampling at low radii, we fixed the minimum bin width at 0.1 in log scale. Given this constraint and the range, we determined the number of bins for each run. After deprojection, each profile is interpolated and extrapolated at the extremities so they all have the same binning. We then calculated the mean 3D-deprojected profile, which is made of 21 bins with a 0.1 width in log scale in the [63.09;6309]~kpc interval. The scatter around the mean profile is the standard deviation over the 100 realisations. The deprojection method was tested on toy models; the mean error of the method is of the order of $6\%$ for the pressure profiles and $5\%$ for the electron density profiles, which is well within the dispersion of the Monte Carlo process.\\

\begin{figure}
   \centering
   \includegraphics[width=\hsize]{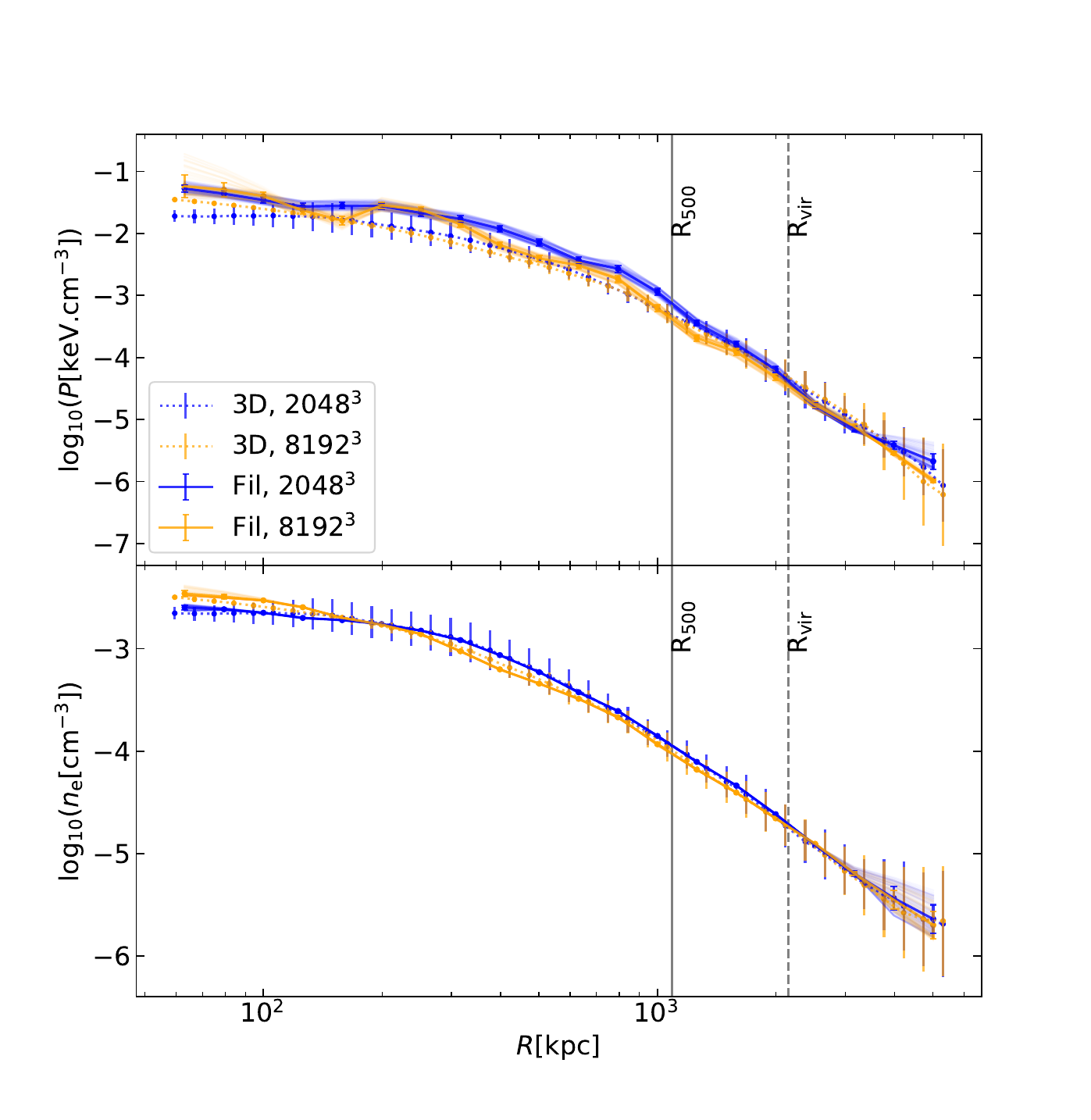}
      \caption{Radial pressure profiles: Orange and blue dotted profiles are the 3D profiles made , respectively, with the high- and low-resolution simulations, and the error bars are the standard deviation. With the same colour code, the solid lines are the 3D-deprojected profiles extracted from the Fil projection. The pale lines are the 100 iterations of the Monte-Carlo smoothing method, and the error bars are the dispersion among these iterations in each bin. The vertical solid line is $R_{500}$, the vertical dashed line is $R_{vir}$. 3D-deprojected profiles are less smooth than their 3D counterparts. The computation of the hydrostatic mass requires the gradients of these quantities, the observed irregularities have a significant impact on the mass estimates.}
         \label{deproj_p_ne_z_hl+3D}
   \end{figure}

\begin{figure*}
        \begin{minipage}[s]{1\textwidth}

            \centering
            \includegraphics[width=.49\textwidth]{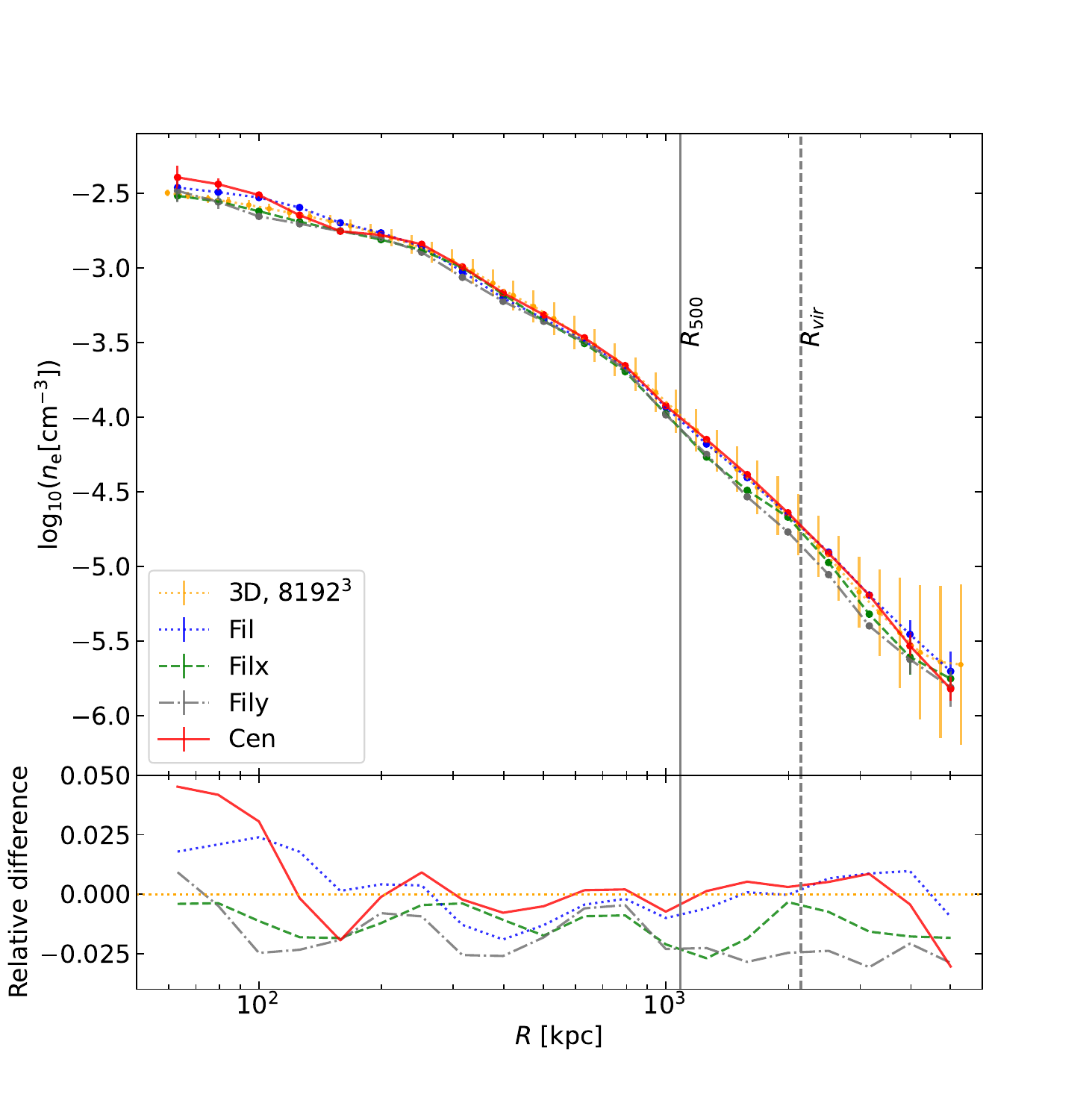}
            \hspace{0.1cm}
            \includegraphics[width=.49\textwidth]{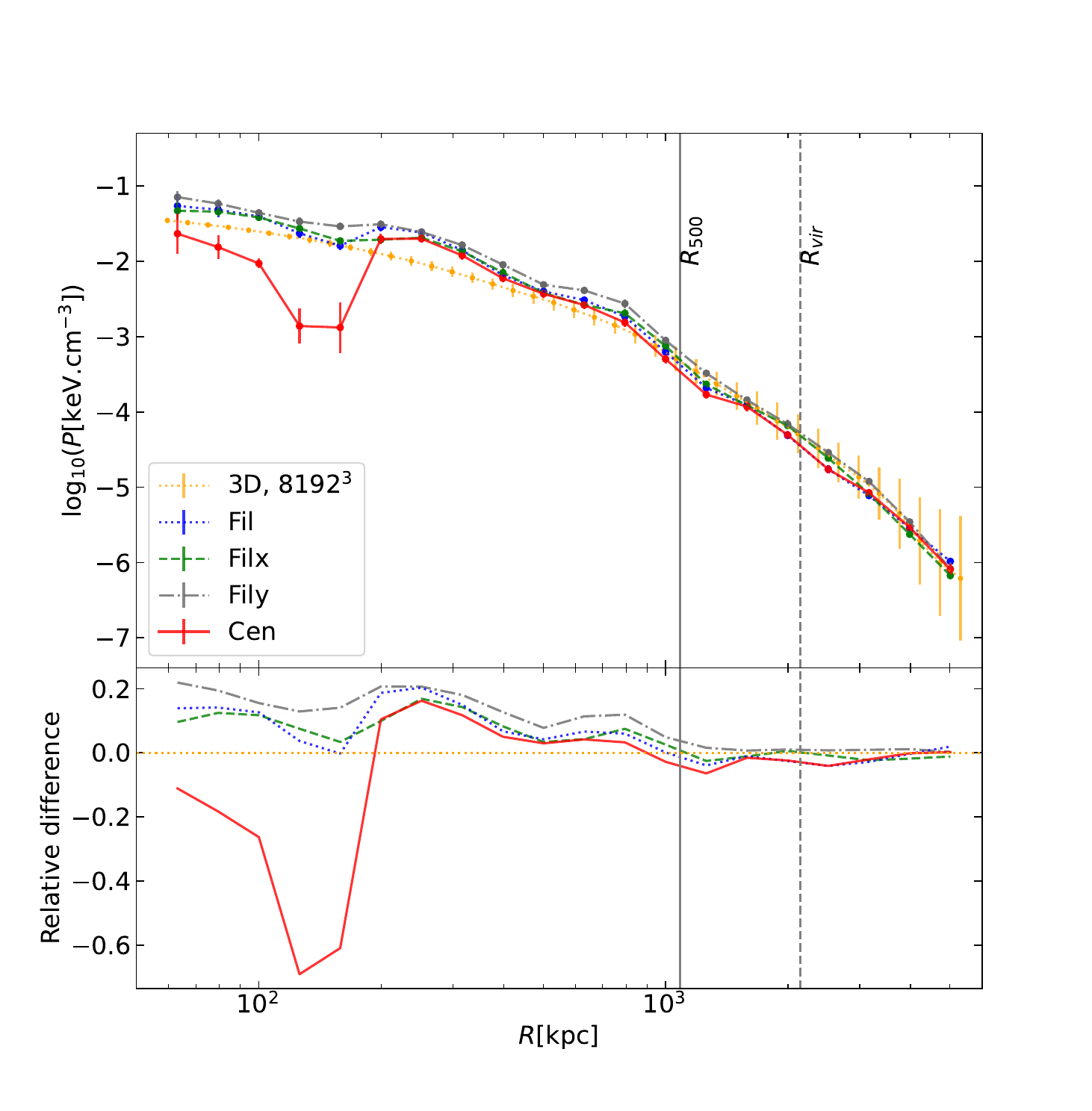}
            \caption{3D radial electron density (top-left) and pressure (top-right) profile from the high-resolution simulation, in orange, compared to 3D-deprojected radial electron density profiles. The blue dotted profile is the Fil projection, the green dashed and grey dash-dotted profiles are the Filx and Fily projections. The red solid profile is the Cen projection. The vertical solid line is $R_{500}$, the vertical dashed line is the virial radius. The bottom panels show the relative difference to the 3D profile for both electron density and pressure profiles. Regarding the correlation between integrated mass in a given direction and profiles intensity, we observe the same trend as for Fig. \ref{proj_profs}. The electron density profiles are not very sensitive to the deprojection process, contrary to the pressure profiles, where the observed features are emphasised and consequently strongly impact the hydrostatic mass derived from these profiles. }
            \label{deproj_profs}
       \end{minipage}
\end{figure*}

We first compared the 3D-deprojected profiles of the low- and high-resolution simulations in Fig. \ref{deproj_p_ne_z_hl+3D}. The top panel shows the pressure and the bottom shows the electron density. The 3D radial profiles are represented by dotted lines for both quantities, whereas the 3D-deprojected profiles are shown with solid lines. The shaded areas around the 3D-deprojected profiles are the 100 realisations of the deprojection process. The high- and low-resolution profiles are, respectively, shown in orange and blue. The vertical solid line is $R_{500}$, and the vertical dashed line gives the virial radius. This figure presents the 3D-deprojected profiles derived from the Fil projection. The 3D-deprojected electron density profiles are very similar to their 3D counterparts for both high- and low-resolution simulations. The 3D-deprojected profiles for the two simulations are similar above 700~kpc. From 250~kpc to 700~kpc, the high-resolution simulation profile is lower, consistently with its 3D counterpart. Closer to the cluster centre, the opposite is true, again consistently with the 3D counterparts. As we discuss more below and show in Fig. \ref{deproj_profs} (left), the 3D-deprojected electron density profiles are consistent with the 3D profile. \\

The 3D-deprojected pressure profile exhibits differences compared to the 3D profiles. For both resolutions, the 3D-deprojected profile is within the error bars of its 3D counterpart from $R_{500}$ to the outskirts. For the high-resolution one, we observe two differences from the 3D profile below  $R_{500}$, the first at 250~kpc and the second at 850~kpc. For the lowest resolution, the 3D-deprojected profile is above its 3D counterpart below  $R_{500}$. We discuss the physical sources of these deviations from the 3D profiles in Section \ref{sec:5}. \\

For the core of the cluster, below 140~kpc the low-resolution simulation 3D-deprojected profile differs from its 3D counterpart, whereas the high-resolution one is higher than its 3D counterpart but within its error bars. The 3D-deprojected profiles are both higher than their 3D counterpart in the core due to the accumulation of small errors during the iterative deprojection process, the error is larger for the low-resolution simulation than for the high one because the maps have fewer pixels. For both the pressure and the electron density 3D-deprojected profiles, the error bars are large in the outskirts due to the extrapolation of the profiles there. \\

Figure \ref{deproj_profs} (left) shows the 3D-deprojected electron density profiles compared to the 3D radial profiles. The bottom panel shows the relative difference to the 3D profile. The differences between the 3D-deprojected profiles and their 3D counterparts are below 5$\%$ and within the error bars of the 3D profile beyond 200~kpc. Figure \ref{deproj_profs} (right) shows the 3D-deprojected pressure profiles compared with the 3D radial profiles. The figure is organised similarly to Fig. \ref{deproj_profs} (left). Contrary to the electron density 3D-deprojected profiles, the pressure is very sensitive to projection effects. Beyond 1~Mpc, the 3D-deprojected profiles are similar to the 3D profile, the deviations are within the error bars. We observe larger differences below $R_{500}$, up to more than 50$\%$. We discuss the projection effects impacting these profiles in Sect. \ref{sec:5}. \\
 
\section{Hydrostatic mass and mass bias}
\label{sec:4}

From the 3D or 3D-deprojected pressure and electron density radial profiles, we can derive the hydrostatic mass and the associated hydrostatic mass bias. In this section, we first present the assumptions made to estimate the mass of clusters and then present and discuss the hydrostatic mass bias estimated at several radii. \\

Assuming that the pressure in the ICM balances the gravitational potential well generated by the cluster total mass, we have a relation between the ICM thermodynamic properties and the cluster mass. Assuming spherical symmetry and only thermal pressure (no turbulent nor magnetic pressure), we obtain the following hydrostatic equilibrium equation: 

\begin{eqnarray}
\frac{\mathrm{d} P_{\mathrm{th}}(r)}{\mathrm{~d} r}=-\frac{G M(<r) \rho(r)}{r^{2}} \; ,
\label{equhydro}
\end{eqnarray}with the thermal pressure $P_{th}$ at a given radius $r$, the density $\rho$, the gravitational constant $G,$ and the total mass $M$ within a sphere of radius $r$.
We also consider a perfect gas to be 

\begin{eqnarray}
    P=\frac{\rho k_{\mathrm{B}} T}{\mu m_{\mathrm{p}}}=n_{\mathrm{e}}k_{\mathrm{B}}T, \qquad   \rho = n_{\mathrm{e}}\mu m_{\mathrm{p}}
\label{perfectgas}
,\end{eqnarray}with the electron density $n_e$, the temperature $T$, the Boltzmann constant $k_B$, the mean molecular weight $\mu$, and the proton mass $m_p$. We can then derive the hydrostatic mass from the pressure and the electron density by writing

\begin{eqnarray}
M_{\mathrm{HE}}(<r)=-\frac{r^{2}}{G \mu m_{\mathrm{p}} n_{\mathrm{e}}(r)} \frac{\mathrm{d} P_{\mathrm{th}}(r)}{\mathrm{d} r}=-\frac{r P_{\mathrm{th}}(r) }{G \mu m_{\mathrm{p}} n_{\mathrm{e}}(r)} \frac{\mathrm{d} \ln P_{\mathrm{th}}(r)}{\mathrm{d} \ln r} \; .
\label{mhep}
\end{eqnarray}Finally, the hydrostatic mass bias $\mathrm({1-b)}$ is defined as the ratio of hydrostatic mass over the total mass within a given radius: 

\begin{eqnarray}
    (1-b)=\frac{M_{\mathrm{HE}}(<r)}{M_{\mathrm {tot}}(<r)}\; .
\label{bias}
\end{eqnarray}

\begin{figure}
   \centering
    \includegraphics[width=\hsize]{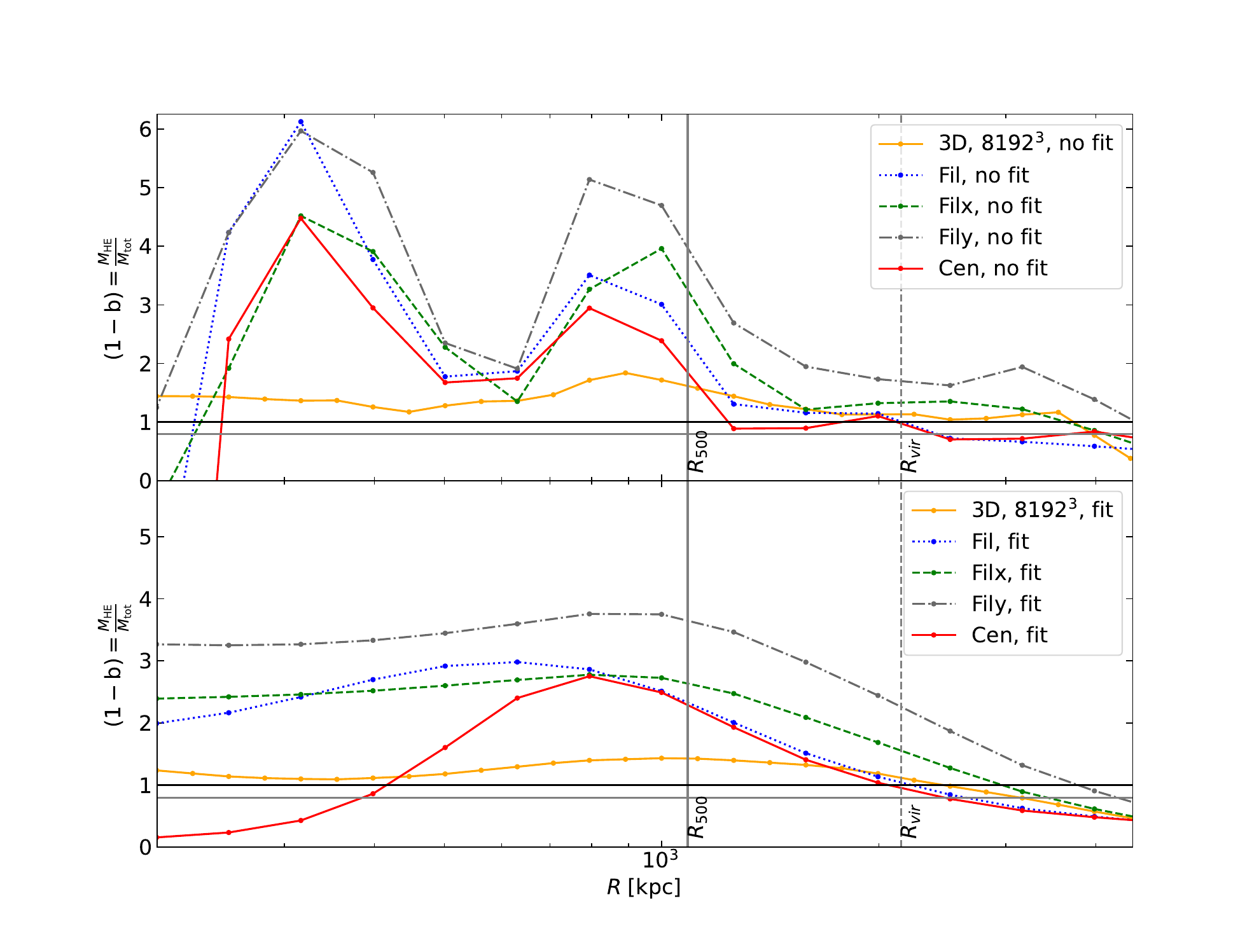}
      \caption{\label{dep_bias} Hydrostatic mass biases estimated at different radii. In orange, the hydrostatic mass bias computed with the 3D radial profiles from the high-resolution simulation. The dashed lines represent the hydrostatic mass biases estimated with 3D-deprojected profiles from several projections. The blue dotted line is the Fil projection, the green dashed and grey dash-dotted lines are the Filx and Fily projections, respectively. The red solid profile is the Cen projection. For both panels, the vertical solid line is $R_{500}$, the vertical dashed line is the virial radius. At around 300 and 850~kpc, the total mass is overestimated up to a factor 6 due to pressure discontinuities observed in Fig. \ref{p maps 4} and \ref{deproj_profs}. The large scatter in the projections at $R_{vir}$ shows the impact of the integrated matter along each direction. The biases obtained from fitted profiles are lower at 300 and 850~kpc but consistent with those from simulation-derived profiles at ${R_{500}}$ and $R_{vir}$. } 
   \end{figure}

 The hydrostatic mass bias at different radii from 200 to 4500~kpc is presented in Fig. \ref{dep_bias}; the colours stand for the same projections as in Figs. \ref{proj_profs} and \ref{deproj_profs}. The hydrostatic mass bias derived from the 3D radial profiles is represented by the solid orange line. If the hydrostatic equilibrium hypothesis were verified, the hydrostatic mass bias would equal one. We observe that 3D hydrostatic mass overestimates the total mass by a factor up to 1.85 at around 850~kpc. At the virial radius, the hydrostatic mass bias is closer to one, meaning the hydrostatic equilibrium hypothesis is valid at this radius. \\

 Using the 3D-deprojected electron density and pressure profiles, the total mass at around 300 and 850~kpc is overestimated by a factor between 2.3 and 6.1, depending on the projection. At the virial radius, the hydrostatic mass biases obtained with the 3D-deprojected profiles ranges between one and two and the scatter is large. The impact of projection effects on the hydrostatic mass is discussed in the following section. \\
 
\section{Projection effects}
\label{sec:5}

In this section, we discuss the impact of projection effects on the 2D radial profiles, 3D-deprojected radial profiles, and the derived hydrostatic mass bias. We distinguish between two main effects: the role of the integrated mass along the LoS in each chosen direction, including the presence of massive objects or substructures that we do not remove, and the signature of small-scale physics in the cluster's core in these directions. \\

Focusing on the 2D radial profiles in Fig. \ref{proj_profs}, the Fily projection has the highest pressure in the cluster core, whereas its electron density counterpart has the lowest profile. On the other hand, the Cen projection shows the lowest pressure profile but the highest electron density profile. We observe the same trend in 3D-deprojected profiles in Fig.\ref{deproj_profs}. \\
    
\begin{figure*}
    \centering
    \includegraphics[width=1\textwidth]{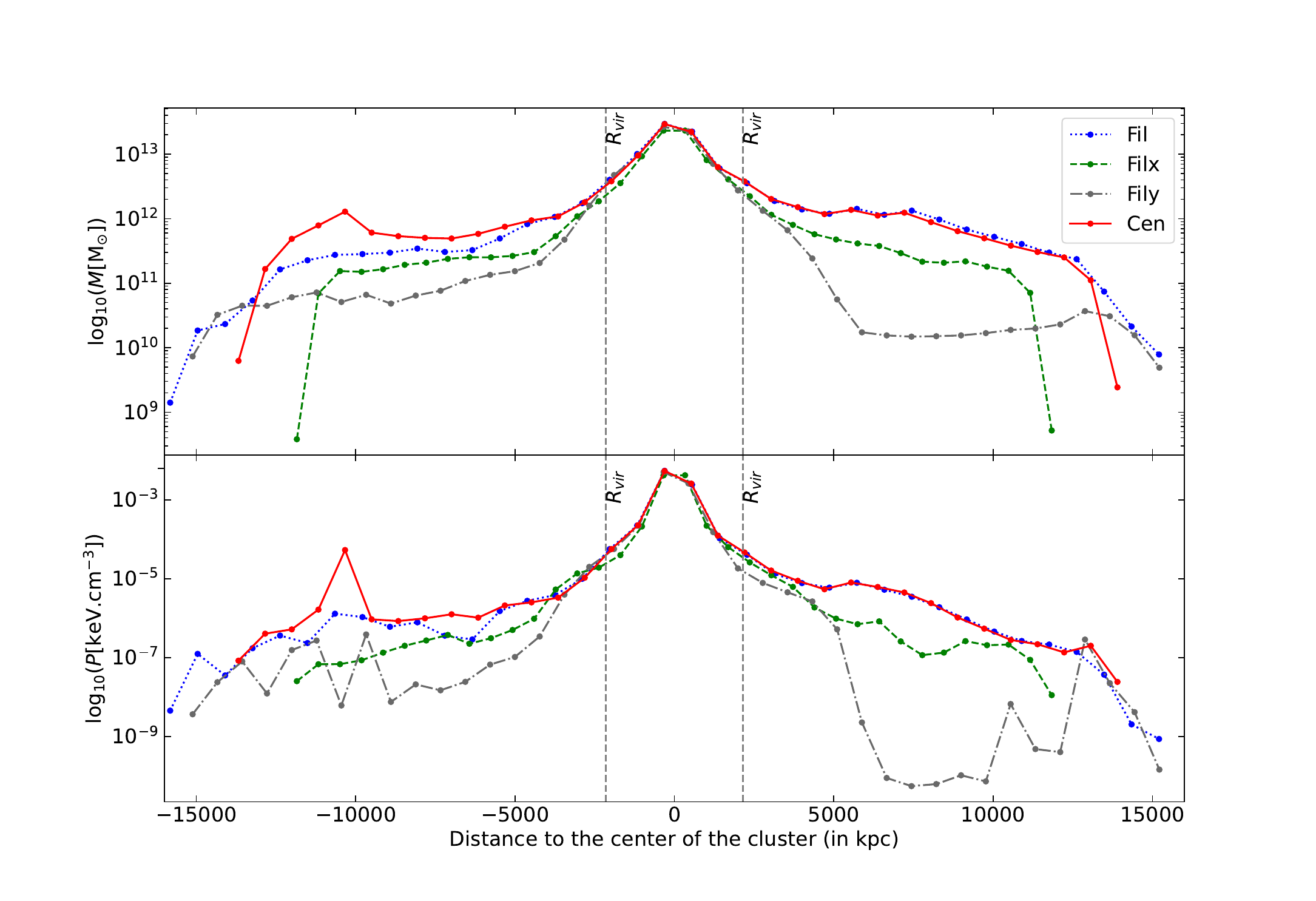}
      \caption{Mass (top) and pressure (bottom) distributions along the LoSs of the four directions. The blue dotted profile is the Fil projection and the green dashed and grey dash-dotted profiles are the Filx and Fily projections, respectively. The solid red profile is the Cen projection. The dashed vertical lines show the virial radius. This figure validates the observed correlation (see Fig.\ref{proj_profs} and \ref{deproj_profs}) between the integrated mass in a given direction and the intensity of electron density and pressure 2D and 3D-deprojected profiles. The pressure drop at 150~kpc in the core of the Cen projection is explained by the presence of a $10^{12}M_{\odot}$ group of galaxies, most probably Canes Venaciti I, along this LoS at 10.3~Mpc from the cluster centre.}
         \label{m_p_dist_los}
\end{figure*}

To explain this trend, we show the gas mass and pressure distribution in the four directions (see Fig. \ref{m_p_dist_los}). The mass is summed in bins of 0.5~Mpc along the LoS in a cylinder of $\mathrm{R_{vir}}$ radius centred on the Virgo cluster position in the plans of the projections. As expected, the mass and pressure peaks are all centred on the Virgo cluster replica. We observe differences in the foreground and background distributions. The Fily projection has the lowest amount of matter and pressure outside the cluster, whereas the Cen projection has the highest. Given that the electron density maps are column densities, this is coherent with what we observe in both Figs. \ref{proj_profs} and \ref{deproj_profs}. The pressure in the pixels of the maps is mass-weighted means of the cells along the LoS. So, if there is mass outside the cluster, the mean 2D radial pressure will be lower than the cluster single mean pressure because the pressure associated with this mass is orders of magnitude lower. This is once again coherent with what we observe in Figs. \ref{proj_profs} and \ref{deproj_profs}. As a result, we can conclude that if there is more mass outside the cluster, the 2D and 3D-deprojected pressure profile is lower and the electron density profile is higher. \\

In Fig. \ref{bias}, we observe how projection effects impact the hydrostatic mass bias estimate. At $R_{vir}$, the hydrostatic mass estimated with the 3D profiles is at one, indicating that the cluster is at the hydrostatic equilibrium at this distance from the centre. So, the dispersion of values at $R_{vir}$ probes the impact of the integrated masses along each LoS. We would intuitively think that the less mass in the foreground and background of the cluster, the more accurate the hydrostatic mass estimate because there is only the contribution of the cluster. We observed the opposite trend. From Eq. \ref{mhep}, we see that the hydrostatic mass depends on the ratio of the pressure over the electron density, so with a higher pressure and a lower electron density the hydrostatic mass is necessarily higher. This explains why the hydrostatic mass estimation for the Fily projection is the highest at every radius, especially at the virial radius. By contrast, the Cen projection has the highest amount of mass and pressure along its LoS, particularly in the foreground as seen in Fig. \ref{m_p_dist_los}. Indeed, the pressure profile is the lowest among the projections while its electron density profile is the highest; as a result, the hydrostatic mass estimation is the lowest almost all the way from the cluster centre to the outskirts. 

Moreover, we observe an important decrease, a reduction by a factor of almost two, on the pressure profile in the cluster's core on the Cen projection 2D radial profile (Fig.\ref{proj_profs} left). This pressure decrease is even stronger on the 3D-deprojected profile (Fig. \ref{deproj_profs}, left). In Fig. \ref{m_p_dist_los}, the mass and pressure distribution along the Cen direction show a secondary peak both in the matter and pressure distribution approximately at 10.3~Mpc from the cluster centre. This massive ($>10^{12} M_{\odot}$) sub-structure is a group of galaxies situated along this peculiar direction. This is most likely the simulated counterpart of the Canes Venaciti I group that is known to be close to the LoS between the Milky Way and the Virgo cluster \citep{karachentsev1966racceannaa,de1975supergalactic,makarov2014structure}.

In Section \ref{sec:2}, we note pressure discontinuities in the pressure maps. There are no discernible signatures of these discontinuities on either 3D pressure or density profiles. The hydrostatic mass bias derived from 3D profiles is slightly impacted by the pressure discontinuity at 850~kpc, and the resulting hydrostatic mass bias is about 1.85 at this specific radius. On the 2D pressure radial profiles displayed in Fig. \ref{proj_profs} (right), we observe variations of the slope at around 250 and 850~kpc. We note an excess pressure at the same radii on the 3D-deprojected profiles (see Fig.\ref{deproj_profs} right). The excess pressure at around 250~kpc is due to the AGN feedback ring visible in Fig. \ref{p maps 4}. The second excess pressure, around 850~kpc, is due to the pressure discontinuity of the gas flowing from the filament into the cluster, as seen in the pressure maps in Fig. \ref{p maps 4}. The deprojection process significantly enhances these pressure discontinuities, and the hydrostatic mass estimated from the deprojected quantities is in turn significantly impacted (see Fig.\ref{dep_bias}). The derived mass is overestimated by a factor between 2.3 and almost 6.1 at these specific radii.

\section{Results and discussion}
\label{sec:6}

\subsection{Simulation convergence}
In the present study, we focused on a simulated replica of the Virgo cluster to showcase the impact of both the integrated mass along each LoS and small-scale physics in the core of the cluster on 2D radial profiles, 3D-deprojected radial profiles, and, consequently, the hydrostatic mass bias. To do so, we made use of the physical quantities directly output from two zoom-in constrained hydrodynamical simulations of Virgo with resolutions of 3D $8192^3$ and 3D $2048^3$. First, we ensured that the simulations had numerically converged and that the results we derived do not depend on the resolution.

Comparing the 3D radial profiles of gas pressure and density derived from the high- (3D $8192^3$) and low- (3D $2048^3$) resolution simulations, we find that they are consistent beyond 150~kpc. In the cluster core, the low-resolution differs from its high-resolution counterpart since it does not fully resolve the small-scale physics. For example, the impact of AGN feedback is smoothed (see Fig. \ref{p maps 2 h l} in Appendix) due to numerical diffusion that is more significant at the 3D $2048^3$ resolution \citep[see Fig. 11 of ][]{teyssier2002cosmological}. To investigate the small-scale physics signatures on the mass bias in the cluster core together with larger-scale projection of matter, we thus based our subsequent study on the high- (3D $8192^3$) resolution simulation of the Virgo replica. We checked that the Virgo masses derived by the two simulations agree. In the left part of Table \ref{tab:mass}, we compare the mass computed at $R_{500}$ obtained in the high- and low-resolution simulations using 3D radial profiles. The mass computed in the low-resolution simulation is 1.8$\%$ lower compared to that in the high-resolution simulation.
It mostly comes from the difference of pressure at $R_{500}$ (see Fig.\ref{deproj_p_ne_z_hl+3D}) due to the numerical diffusion of the accretion shock at $850kpc$. Nevertheless, the difference is minor, and the two simulations are in excellent agreement. As a matter of fact, the lower resolution simulation has fully reached convergence beyond the core region ($>150$~kpc), so usual physical quantities computed at $R_{500}$ or $R_{vir}$ are fully coherent with the higher resolution simulations and can thus be reliably used. 

\begin{table*}[]
    \centering
    \caption{Mass of the Virgo cluster in $10^{14}M_{\odot}$ at $R_{500}$ (top) and at $R_{vir}$ (bottom).}
    \renewcommand{\arraystretch}{1.2}
    \centering
    \begin{tabular}{ c c c c c c }
        \hline\hline
         \multicolumn{1}{c}{} & \multicolumn{5}{c}{Virgo masses at $R_{500}$ (\boldmath$10^{14} M_{\odot}$)} \\
         \hline
         \multirow{8}{*}{Sim.} & \multicolumn{5}{c }{$M_{3D,500}=3.36$} \\
         \cline{2-6}
         {} & \multicolumn{5}{c }{$M_{HE}$ (Hydrostatic mas) }\\
          & & Sim-derived & Fit & + $P_{nth}$ & + $P_{nth}$+fit \\
         \cline{2-6}
       
       {} & \multicolumn{1}{c}{3D $8192^3$}  & 5.76 & 5.48 & 5.95 & 5.59 \\ 
       {} & \multicolumn{1}{c}{3D $2048^3$}  & 5.66 & 5.26 & 5.84 & 5.36 \\
       {} & \multicolumn{1}{c}{Fil} & 9.46 & 7.90 & - & -  \\
       {} & \multicolumn{1}{c}{Filx} & 12.45 & 8.58 & - & -  \\
       {} & \multicolumn{1}{c}{Fily} & 14.76 & 11.79 & - & -  \\
       {} & \multicolumn{1}{c}{Cen} & 7.51 & 7.84 & - & - \\
       \hline
       \multirow{1}{*}{Obs.} & \multicolumn{5}{c }{$0.83\pm0.01$ (from X-rays using clmass model) \citep{simionescu2017witnessing} } \\
       \hline
       
        \multicolumn{6}{c}{} \\
        \hline \hline
        \multicolumn{6}{c}{Virgo masses at $R_{vir}$ (\boldmath$10^{14} M_{\odot}$)} \\
        \hline
        \multirow{1}{*}{Sim.} & \multicolumn{5}{c}{$M_{3D,vir}=6.31$} \\
       \hline  
       \multirow{5}{*}{Obs.} & \multicolumn{5}{c}{6 \citep{de1960apparent}} \\
       {}  & \multicolumn{5}{c}{2.7-8.9 \citep{karachentsev2010observed}} \\
       {} & \multicolumn{5}{c}{8.0$\pm$2.3 \citep{karachentsev2014infall}} \\
       {}  & \multicolumn{5}{c}{7 \citep[][updated values as given in EDD database \tablefootmark{a} ]{tully2015galaxy}} \\ 
       {}  & \multicolumn{5}{c}{6.3 $\pm$ 0.9 \citep{kashibadze2020structure}} \\
       \hline
    \end{tabular} 
    \label{tab:mass}
    \tablefoot{On the first row of each table, we give the total of the Virgo replica (i.e. the sum of DM and baryons in a sphere of the given radius). In the upper table, in the simulation (Sim.) sub-panel, we present the hydrostatic mass derived from 3D profiles in the high- (3D $8192^3$) and low- (3D $2048^3$) resolution simulations and from projections (Fil, Filx, Fily, and Cen). We present the hydrostatic mass computed from simulation-derived profiles (Sim-derived) and from fitted electron density and pressure profiles (Fit). For the 3D profiles, we also show the hydrostatic mass derived when adding non-thermal pressure while fitting the profiles, including the $\alpha$ parameter (+$P_{nth}$+fit) or not (+$P_{nth}$). In the lower sub-panel (Obs.) of each table, we present the observed masses of the Virgo cluster estimated from X-rays at $R_{500}$ (top) and from galaxies velocity dispersion at $R_{vir}$ (bottom).\tablefoottext{a}{\url{https://edd.ifa.hawaii.edu/}}}
\end{table*}

\subsection{Radial profile choice}

A notable difference in the approach followed in our study with respect to other studies is the definition of the profiles. More specifically, in contrast with the majority of studies using fitted profiles for the pressure, electron density, and temperature \citep[e.g. ][]{kay2012sunyaev,martizzi2016mass,gupta2017sze,henson2016impact,pearce2020hydrostatic,ansarifard2020three,barnes2021characterizing,gianfagna2021exploring}, we instead estimated the hydrostatic mass from the actual radial profiles of pressure and electron density. This choice was motivated by our interest in the detailed impact of physical effects on the mass determination. In the following, we discuss the impact of fitting the radial profiles on the hydrostatic mass as compared to that derived from profiles directly obtained from the simulations.

\begin{figure}
   \centering
    \includegraphics[trim=20 10 20 10, width=\hsize]{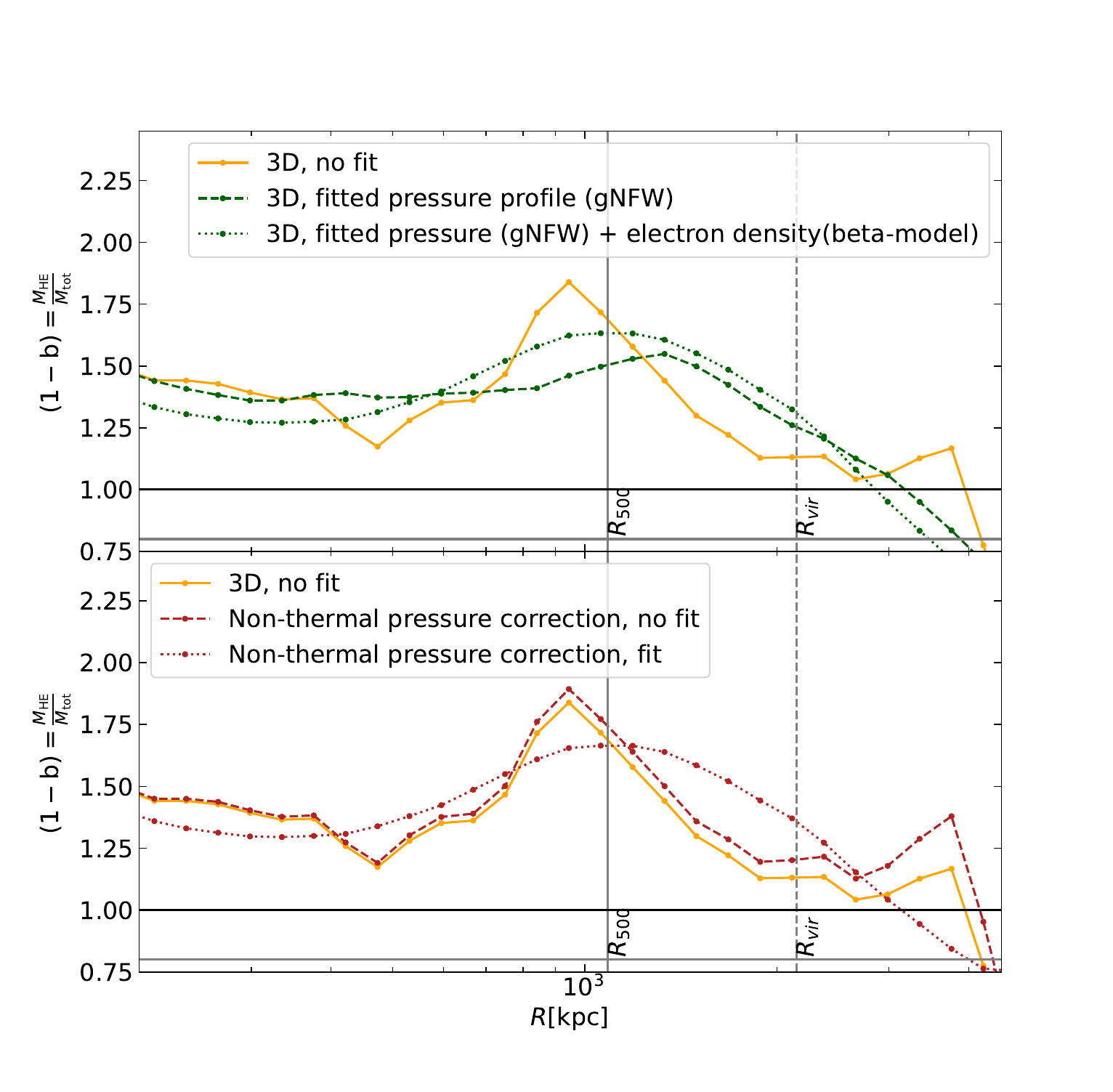}
      \caption{\label{bias_fit} Hydrostatic mass biases estimated at different radii. For both panels, the orange line is the hydrostatic mass bias computed with the 3D radial profiles from the high-resolution simulation considering only thermal pressure. The vertical solid line is $R_{500}$ and the vertical dashed line is the virial radius. Top: Dark green dashed line is the bias, while only fitting the pressure to gNFW model, and the dark green dotted one is the bias while fitting both pressure and electron density profiles. Bottom: Dark red lines stand for the hydrostatic mass bias with non-thermal pressure correction, with and without fitting the radial profiles to universal models (pressure, electron density and the $\alpha$ parameter), respectively, for the dotted and the dashed line styles. The bias deduced from both fitted pressure and electron density profiles has a similar trend to that obtained from simulation-derived profiles: the values are close at $R_{500}$. Virgo being unrelaxed, adding non-thermal pressure increases the mass even more. The comparison between the bias deduced from fitted and simulation-derived profiles leads to the same conclusions as the top panel.} 
\end{figure}

Starting with the Virgo-replica mass, we observe that the mass derived from the 3D simulation-derived profiles is 4.9$\%$ lower than that from fitted profiles (Table \ref{tab:mass}). In Sect. \ref{sub6.virgoenv}, we discuss how this small difference is significantly increased depending on the LoS.\\
Focusing on the hydrostatic mass bias, we discuss the impact of fitting the radial profiles by contrast to using the 3D simulation-derived profiles. The top panel of Fig. \ref{bias_fit} shows the hydrostatic mass bias, similarly to Fig.\ref{dep_bias}. The bias estimated from the 3D profiles without fitting is the orange solid line, the bias estimated while fitting the pressure profile to the gNFW model is the dark green dashed line, and the bias estimated while fitting the pressure and the electron density to the $\beta$ model is the dark green dotted line. By fitting the pressure profile, the dip at 480~kpc is increased by $20\%$, and the peak at 950~kpc is reduced by $20\%$. Within $R_{500}$, the hydrostatic mass bias does not depend much on the radius compared to the bias estimated without fitting the radial profiles. Then, from 1.2~Mpc to 3~Mpc, the bias computed from the fitted pressure profile is larger than that computed without fitting any profile. Finally, the two biases, with or without fitted radial profiles, are close to one between 2.5 and 3~Mpc. \\
The bias estimated from the fitted pressure and electron density profiles shows a smoother behaviour than that estimated from 3D simulation-derived profiles. The general trend is the same: a dip at 400~kpc and a peak at $R_{500}$. The three biases have the same value at about 2.5~Mpc. In the case of the bias estimated from the 3D profiles, we note that fitting the profiles smooths the small variations. Still, the estimated bias at $R_{500}$ and 2.5~Mpc are quite consistent regardless of whether a fitted profile or a simulation-based one is used. As a matter of fact, 3D pressure and electron density profiles are already very smooth, so fitting those profiles did not change their shape and the derived mass bias very much.

\subsection{Virgo environment and projection effects}
\label{sub6.virgoenv}
One of our goals is to investigate the projection effects on Virgo properties, that is, the impact of mass integration on the mass and
mass-bias estimates in given directions. In this context, our use of a constrained simulation, providing us with the Virgo replica, has the unique advantage of actually mimicking the cluster's large-scale environment. We took advantage of this by building projected $(10R_{\mathrm{vir}}\times10R_{\mathrm{vir}}$) maps along $15R_{\mathrm{vir}}$ following four directions. These large maps contrast with previous studies (e.g. \citeauthor{ameglio2007joint} \citeyear{ameglio2007joint} built projected $2R_{\mathrm{vir}}\times2R_{\mathrm{vir}}$ X-rays and SZ maps along $6R_{\mathrm{vir}}$).

Moreover, we used simulation outputs of pressure and density rather than reconstructed profiles from mock tSZ and X-ray \citep[e.g.][]{ameglio2007joint,ameglio2009reconstructing,barnes2021characterizing,zuhone2023effects} or lensing \citep[e.g.][]{meneghetti2010weighing,munoz2023galaxy} observations. As a consequence, we addressed the impact of all the substructures and massive objects along the LoS 
that are usually subtracted from actual data (e.g. X-rays) and in the above-cited works. In a forthcoming work aimed at providing a comparison with observations, we will use mock tSZ and X-ray emission maps and follow the usual observational approach of excluding clumps and substructures. \\

In the constrained simulation of Virgo, we observe a group of galaxies in our Cen projection, which does not affect the electron density map much, but has an important impact on the pressure map. Although it is massive, the group along the LoS has a much lower pressure than that in the cluster. Thus, the mass-weighted mean pressure appears to be lower when considering both the group and the cluster rather than the cluster alone. Consequently, we observe a decrease by more than 60$\%$ in the Cen projection compared to other projections. Such matter projection along the LoS was also observed by \citet[Figure 12]{ameglio2007joint} who found a gas clump in their z projection. However, given that its gas pressure is comparable to that of the main cluster and given their use of X-ray and tSZ maps, they found different contributions with respect to ours. Therefore, the intensity of the impact of massive objects along the LoS depends on the studied projected quantities. \\

Projecting along the main filament connected to Virgo and orthogonally to it, we show that when we integrate more mass along the main filament, the pressure maps are thus decreased, and the electron density maps are increased. This results in a smaller hydrostatic mass.  \cite{gouin2021shape} showed that cluster shape and connectivity are tightly linked. Hence, projecting given filament orientations can be related, to some extent, to projecting along semi-principal axes. \cite{barnes2021characterizing} studied the impact of clusters' triaxiality on the hydrostatic mass estimation for a large sample of clusters. They conclude that the orientation does not significantly impact the average hydrostatic mass estimation and that the scatter in individual mass estimation dominates the orientation choice. However, it is worth noting that a direct comparison between our results and those of \cite{barnes2021characterizing} is not possible given our different scales of interest and our different methodologies. In their work, they used mock X-ray maps of $3R_{\mathrm{500}}\times3R_{\mathrm{500}}$ generated from a sphere of radius $5R_{\mathrm{200}}$; they did not consider a background in the deprojection and they used the L1 regularisation method \citep{ameglio2007joint}. Whereas, we used $10R_{\mathrm{vir}}\times10R_{\mathrm{vir}}$ maps of pressure and electron density projected along $15R_{\mathrm{vir}}$, and we took into account the background and used a geometrical deprojection method. \\

We focused on the impact of the projection effect on the estimated cluster mass and show, in Table \ref{tab:mass}, the Virgo total and hydrostatic masses at $R_{500}$ derived from the different projections defined in Sect. \ref{subsec:2.3}. We observe a significant impact on the total mass derived from deprojected profiles compared to that obtained from the raw 3D profiles. The mass is found to be 15.5$\%$ lower for the Fil projection, 30.3$\%$ lower for the Filx projection, 19.2$\%$ lower for the Fily projection, and 5.5$\%$ higher for the Cen projection. Due to the smoothing of discontinuities in the fitted profiles\footnote{For the Cen projection, the 3D-deprojected pressure profile is strongly disturbed by the presence of the group of galaxies along the LoS, inducing an unreliable fitting.}, masses estimated from these profiles are also significantly lower for Fil, Filx, and Fily than that derived from the raw 3D profiles. We can compare our mass estimations of the Virgo replica to some masses derived from observations in the X-rays \citep[][left]{simionescu2017witnessing} and in the optical \citep[][right]{de1960apparent,karachentsev2010observed,karachentsev2014infall,tully2015galaxy,kashibadze2020structure}. The Virgo replica total mass, computed at $R_{vir}$, $6.31\times10^{14}M_{\odot}$, is in rather good agreement with the observed masses that range from $6\times10^{14}M_{\odot}$ to $8\times10^{14}M_{\odot}$. In the bottom part of Table \ref{tab:mass}, we present observed Virgo mass at $R_{500}$ \citep[][left]{simionescu2017witnessing} estimated from X-rays observations. It is important to note that Virgo is the closest cluster to us; consequently, and unlike the majority of galaxy clusters, estimating the mass of Virgo from tSZ \citep{planck2016virgo} or X-rays \citep{young2002chandra,urban2011x,simionescu2015uniform,simionescu2017witnessing} observations is very challenging given that Virgo is very extended in the sky.

\subsection{Mass bias}
We now turn to the discussion of the mass bias results. We first emphasise the impact of the projection effects, combined with specific signatures of physical process in the core at particular distances from the centre Virgo-replica, on the mass biases derived from 3D-deprojected pressure and electron density profiles (see Fig. \ref{dep_bias}). We find that at 300~kpc, the biases from the raw simulation-derived profiles can reach large values between 4.4 and 6.1 (between 0.4 and 3.3 when the profiles are smoothed). At 850~kpc, biases range between 2.9 and 5.1 (between 2.7 and 3.8 for the smoothed profiles). For the Cen projection, the pressure drop in the core implies a lower fitted pressure profile and a smaller bias estimation inside 850~kpc. At $R_{500}$ and beyond, the detailed core physics does not impact the cluster properties and the biases range between 1.8 and 4 (between 2.2 and 3.7 using smoothed profiles) at $R_{500}$ and between 1.1 and 1.7 (between 1 and 2.3 using smoothed profiles) at $R_{vir}$.

\begin{figure}
    \centering
    \includegraphics[width=\hsize]{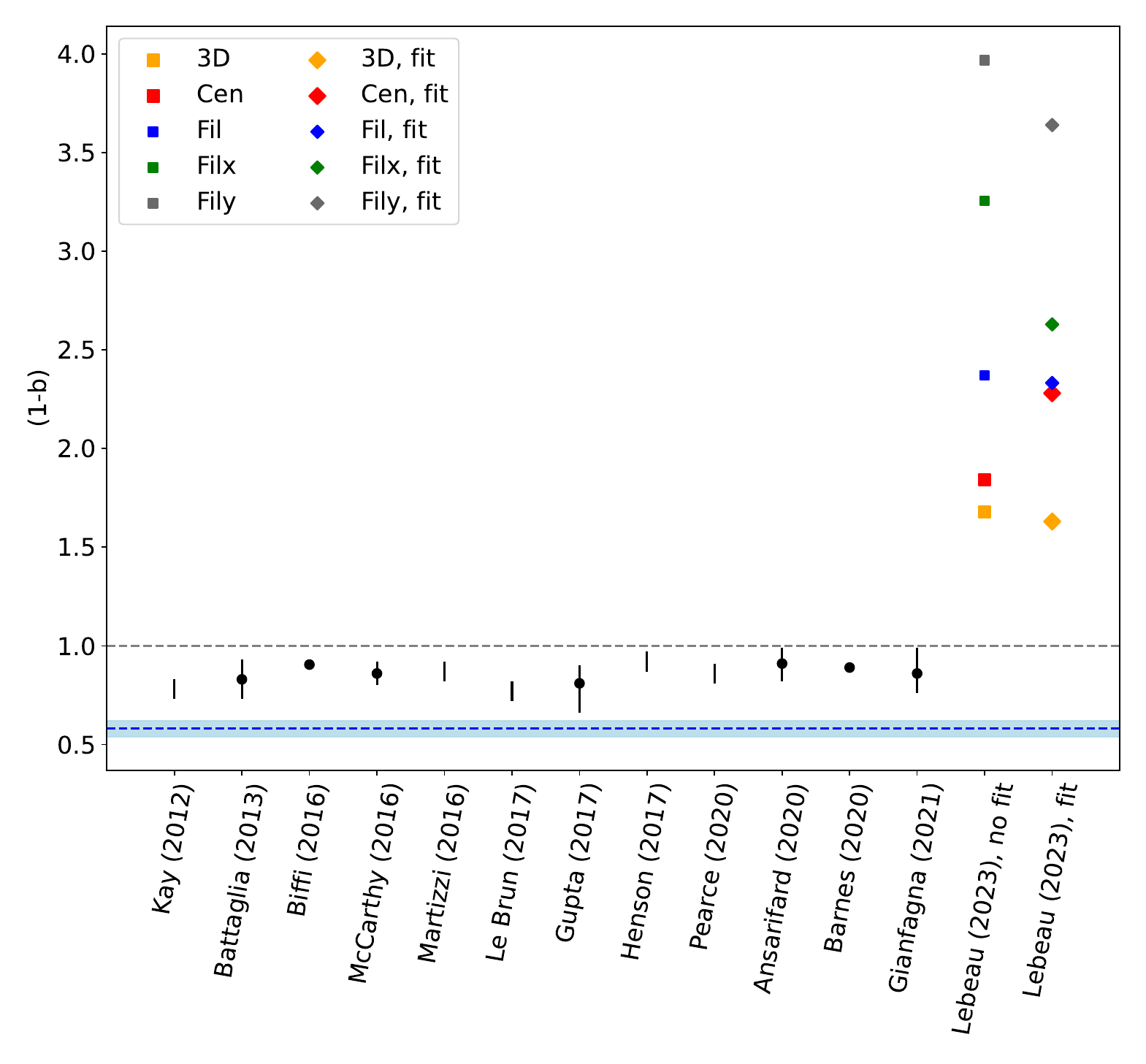}
    \caption{Comparison of average hydrostatic mass biases at $R_{500}$ from samples of clusters in random non-constrained simulations with the values obtained using 3D or the 3D-deprojected profiles from the constrained simulations of the Virgo replica. The constrained simulation includes the same sub-grid physics as the random simulation with the addition of the quasar mode for the AGN feedback. We used the same colours for the biases obtained in this work. Virgo being an unrelaxed cluster, the bias obtained from the 3D profiles is larger than the average biases from the random simulations. The projection effects emphasise the variations in the profiles and induce very large biases.}
    \label{litterature_biases}
\end{figure}

In Fig. \ref{litterature_biases}, we display our values of the hydrostatic mass bias at $R_{500}$ for the Virgo replica obtained with the 3D and the 3D-deprojected profiles compared to the values in the literature \citep[e.g. ][]{kay2012sunyaev,battaglia2013cluster,biffi2016nature,mccarthy2016bahamas,martizzi2016mass,le2017scatter,gupta2017sze,henson2016impact,pearce2020hydrostatic,ansarifard2020three,barnes2021characterizing,gianfagna2021exploring}. In doing so, it was important to note and bear in mind the specificities of our study. Namely, we did not follow the observational approach (raw simulation-driven or deprojected profiles as compared to X-ray,tSZ or lensing-derived ones); we used raw rather than fitted profiles; we focused on a single, non-relaxed cluster of a given mass rather than deriving average quantities for a full sample of different masses; and, finally, we used a constrained simulation including the two AGN-feedback processes and intended to reproduce the cluster rather than a random initial-condition simulation.\\

The values of mass bias from the literature in Fig. \ref{litterature_biases} are means (or medians for \citeauthor{biffi2016nature} \citeyear{biffi2016nature} and \citeauthor{gianfagna2021exploring} \citeyear{gianfagna2021exploring}) over large samples of simulated clusters. They are distributed around 0.8, with a mean value of 0.855 and a standard deviation of 0.047, and differ from the value needed to reconcile tSZ cluster counts with the CMB \citep{ade2016planck}. It is worth noting that the average/median biases are estimated from cluster samples without distinguishing between relaxed and unrelaxed clusters. A larger scatter around the median bias might be expected for disturbed clusters such as Virgo. For example, \citet{gianfagna2021exploring} showed that values range from 0.65 to 1 for disturbed clusters and 0.72 to 0.94 for relaxed ones. Similarly, an increased scatter was found in \citet{ansarifard2020three}, which also found that the irregular clusters' median bias presents higher departure from hydrostatic equilibrium. Not all studies reach such conclusions \cite[e.g.][]{biffi2016nature}. This can be due to differences in the simulation used, notably the implementation of AGN/SN feedback impacting the cluster cores and thus the hydrostatic mass estimation at $R_{500}$. They could also be due to the samples of selected clusters and their classifications in perturbed or unperturbed states.

In contrast with the median/average values in Fig. \ref{litterature_biases} or those accounting for the increased scatter due to the dynamical state, the biases we estimated at $R_{500}$ for the Virgo replica are significantly larger. Depending on the direction of observation, they range from 1.8 to 4 (2.2 and 3.7) when using 3D-deprojected (fitted) profiles. The important spread in Virgo's bias values illustrates the importance of the impact of projection along some directions on the mass bias estimate. In the direction similar to ours, the bias is 1.8 and 2.3 depending on whether the 3D-deprojected profiles are fitted or not. It is smaller than the other quoted values but still $\sim5-7\sigma$ away from the median absolute deviation interval at $R_{500}$ for disturbed clusters in \cite{gianfagna2021exploring}. Such large values of the hydrostatic mass bias are not uncommon for individual simulated clusters. For instance, \citet{kay2012sunyaev} showed that the biases of individual clusters estimated from 3D fitted radial profiles range between 0.31 and 2.2, comparable to the value of 1.6 we obtain from the fitted profile of Virgo replica. More recently, \citet{ansarifard2020three} estimated the bias from direct 3D simulation outputs and showed that some clusters have a bias of 1.4 at $R_{500}$, similar to the bias of 1.6 estimated from 3D fitted profiles of the Virgo replica. Finally, following a different approach based on scaling relations, \citet{biffi2016nature} also showed that the bias of individual clusters is widely spread around the median bias, reaching values above 1.5. These comparisons confirm that our Virgo replica shares the same hydrostatic mass bias value as other random clusters, although such clusters are uncommon. This is in agreement with the fact that the Virgo cluster is an unlikely cluster of galaxies \citep{sorce2019virgo}.

\subsection{Contribution from non-thermal processes}

As discussed above, the Virgo cluster is not relaxed, and the complex physics effects (AGN feedback and matter infall from the filament) are observed in the cluster core within $R_{500}$. The hydrostatic equilibrium assumption is invalid, and we expect that the derived hydrostatic mass and bias differ from the expected values for relaxed clusters. 
When estimating the hydrostatic mass, only the pressure due to gravitational heating was accounted for. However, magnetic fields \citep[e.g.][]{pellissier2023rhapsody}, cosmic rays \citep[e.g.][]{boss2023crescendo}, or turbulence can provide additional pressure support \citep[e.g.][]{nelson2014hydrodynamic,pearce2020hydrostatic}. 
To estimate its specific contribution to the case of the Virgo replica, we modelled the non-thermal pressure following \cite{gianfagna2021exploring}. \\
\indent Figure \ref{bias_fit} (bottom) shows the hydrostatic mass bias with the non-thermal pressure correction in red compared to the bias considering only thermal pressure. The red dotted and dashed lines, respectively, show the hydrostatic mass bias with and without fitting the radial profiles to the universal models. The non-thermal correction adds pressure and eventually gives a higher hydrostatic mass bias. As observed by \cite{nelson2014hydrodynamic} and \citep{pearce2020hydrostatic}, the non-thermal pressure support increases with radius and so the hydrostatic mass bias. In \cite{gianfagna2021exploring}, the mean bias is below one, so adding pressure reduces the bias. For the case of the Virgo replica, the total mass considering only thermal pressure is overestimated, adding the non-thermal contribution increases the mass further. We then fitted the pressure, electron density, and $\alpha$ \citep[defined as the ratio of the non-thermal pressure over the total pressure in][]{pearce2020hydrostatic} profiles; the bias computed with the fitted profiles is shown as the dark red dotted line. It has the same variations and values as the bias estimated from fitted pressure and electron density profiles shown in dark green dotted line in the top panel of Fig. \ref{bias_fit}. Finally, the three biases presented in the bottom panel are around 1.15 at 2.5~Mpc, which is the same as the biases in the top panel. In Table \ref{tab:mass}, we give the Virgo replica masses at $R_{500}$ calculated while adding non-thermal pressure while fitting the radial including the $\alpha$ parameter  (+$P_{nth}$+fit) or not (+$P_{nth}$). In the high-resolution simulation, the mass calculated from simulation-derived profiles while adding non-thermal pressure is 3.1$\%$ higher than when considering only thermal pressure. Using the fitted profiles, it is 1.9$\%$ higher. The difference is thus not significant.

\section{Conclusion}
\label{sec:7}

In this work, we studied the impact of projection effects on the hydrostatic mass estimation of the simulated Virgo cluster replica. We based our results on an $8192^3$ effective resolution hydrodynamical zoom-in simulation of Virgo  from a larger constrained simulation. We computed the mass under the hydrostatic assumption in different cases using: i) 3D radial profiles of pressure, electron density, and temperature from selected ICM cells (simulation-derived profiles); ii) 2D radial pressure and electron density profiles in concentric annuli from maps projected in four different directions (Cen, Fil, Filx, and Fily); and iii) 3D-deprojected profiles using a model-free geometric deprojection technique.\\

\indent In addition, we compared our work to other projection-effect studies. We also discuss the impact of using fitted, hence smoother, radial profiles rather than the simulation-derived ones. Moreover, we compared the total and hydrostatic mass of the Virgo replica estimated either from 3D and 3D-deprojected radial profiles to estimations of the real observed Virgo mass. We then compared our values of the hydrostatic mass bias, at $R_{500}$, for the Virgo replica obtained with the
3D and the 3D-deprojected profiles to the values obtained for samples of clusters of different masses extracted from random, that is non-constrained, cosmological simulations. Finally, we considered the addition of a non-thermal pressure term and its impact on the Virgo replica mass determination. \\

We found that the mass distributed along the LoS, in the foreground or background, impacts the 2D radial and 3D-deprojected profiles. For the electron density, the more mass, the higher the overall profile. For the pressure, the more mass, the lower the profile. This result might seem counter-intuitive given that we expect higher pressure intensity with more pressure along the LoS. This would be the case for SZ maps because the pressure is integrated along the LoS. However, since we used mass-weighted pressure maps, pressure clumps outside the cluster, that are indeed orders of magnitude lower, will lower the mass-weighted mean pressure. The constrained simulation of Virgo shows a group of galaxies, most likely the Canes Venaciti I replica along the Cen direction,  which is close to our real line of sight. Its presence along the LoS induces an important pressure decrease in the core of the 3D-deprojected pressure profile. The Fil projection, that is that along the main filament, contains the largest mass distribution in the foreground and background, whereas the Fily projection, that is that perpendicular to the main filament with a rotation around the y-axis, contains the least. At $R_{vir}$, the bias derived from 3D profiles is one, indicating that the hydrostatic equilibrium is reached at this radius. The scatter of the hydrostatic mass bias values estimated from the different projections, at this distance, shows the impact of the integrated mass along the different LoSs. The Fily projection has the least integrated mass along its LoS, leading to the highest pressure and lowest electron density, it thus has the highest hydrostatic mass estimation among the projections. It is the opposite for the Fil projection. Those two projection effects, namely the presence of a group of galaxies along the LoS of a given projection and the impact of the local environment around clusters on projected quantities, can affect any cluster mass estimation. \\

Virgo is known to be an unrelaxed cluster. For this Virgo replica, we find that the hydrostatic equilibrium assumption is invalid within $R_{500}$. We indeed observe two significant pressure discontinuities in the inner part of the cluster. The first, at 300~kpc, is due to AGN feedback of central galaxy M87. The second, at 850~kpc, is due to mass infall from a filament connected to Virgo. As a result, the hydrostatic masses estimated from the 3D profiles at these specific radii are overestimated by factors of 1.4 and 1.85. Once again, projection effects impact the mass determination. The pressure discontinuities are even more emphasised by the deprojection method, leading to a very significant overestimation of the cluster mass, up to a factor of 6.1, in the Fil projection at 300~kpc. \\

This study of the Virgo cluster replica reveals the complex physics at play in this object and shows the different types of projection effects. It is a case study of a very specific unrelaxed cluster considered as a first step of a larger project within the LOCALIZATION\footnote{\url{https://localization.ias.universite-paris-saclay.fr/}} collaboration, in which we will study the contributions of different sources of bias on the mass determination for a large sample of galaxy clusters extracted from a constrained hydrodynamical simulation of the local Universe. We will then be able to compare the results to those of random simulations and actual observations. 

\begin{acknowledgements}
    The authors thank the referee for helping to improve this article. This work was supported by the grant agreements ANR-21-CE31-0019 / 490702358 from the French Agence Nationale de la Recherche / DFG for the LOCALIZATION project. The authors were also supported by funding of the ByoPiC project from the European Research Council (ERC) under the European Union’s Horizon 2020 research and innovation program grant agreement ERC-2015-AdG 695561 (ByoPiC, https://byopic.eu). The authors thank the very useful comments and discussions with all the members of this project. They thank the Centre for Advanced Studies (CAS) of LMU Munich for hosting the collaborators of the LOCALIZATION project for a week-long workshop. 
    The authors gratefully acknowledge the Gauss Centre for Supercomputing e.V. (www.gauss-centre.eu) for providing computing time on the GCS Supercomputers SuperMUC at LRZ Munich. 
    We finally thank Florent Renaud for sharing the rdramses RAMSES data reduction code.
\end{acknowledgements}

\bibliographystyle{aa} 
\bibliography{bibliography}

\appendix
\onecolumn

\section{Deprojection algorithm}\label{appendix_algo}

\begin{figure}
   \centering
    \includegraphics[width=0.4\textwidth]{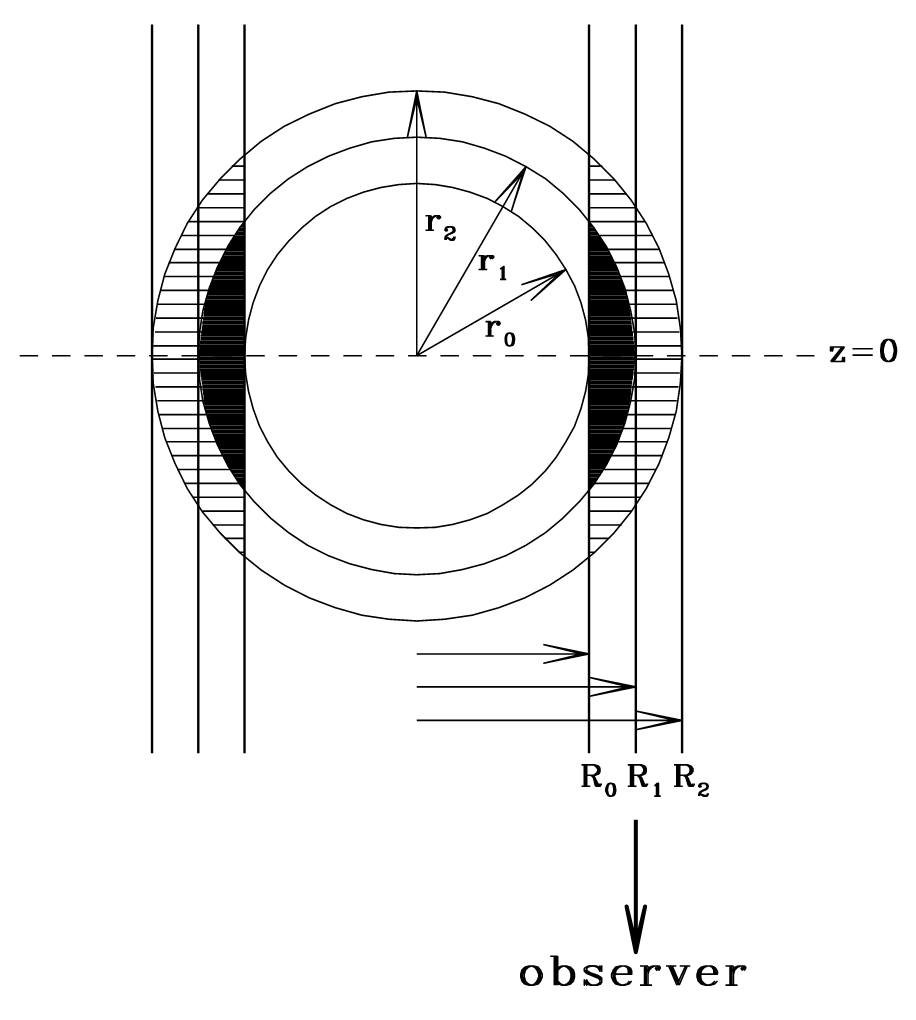}
      \caption{Illustration of a geometrical deprojection algorithm. The plane of the sky is perpendicular to the page. Radii $r_0$, $r_1$, and $r_2$ are 3D quantities and define spherical shells. Radii $R_0$, $R_1$, and $R_2$ are projected quantities referring to cylindrical shells along the LoS; these correspond to circular annuli on the plane of the sky. The figure is extracted from \cite{mclaughlin1999efficiency}.}
         \label{dep_scheme}
   \end{figure}

The deprojection algorithm detailed in this appendix is inspired mainly by the method described in the appendix of \cite{mclaughlin1999efficiency}. To describe the method, we take a simple example, as shown in Fig. \ref{dep_scheme} \citep[from][]{mclaughlin1999efficiency}. We assume that the cluster is a perfect sphere, we divide the sphere into a central core and two spherical shells; the radii of the sphere and the two shells are, respectively, $r_0$, $r_1,$ and $r_2$. This sphere is intersected by cylindrical annuli of radii $R_0$,$R_1$ and $R_2$; for simplicity we have $r_i=R_i$ for all $i$. The volume made of the intersection of a cylindrical annulus and a spherical shell is the following:

 \begin{equation}
V_{\mathrm{int}}\left(r_{j-1}, r_{j} ; R_{i-1}, R_{i}\right)=\frac{4 \pi}{3}\left[\left(r_{j}^{2}-R_{i-1}^{2}\right)^{3 / 2}-\left(r_{j}^{2}-R_{i}^{2}\right)^{3 / 2}+\left(r_{j-1}^{2}-R_{i}^{2}\right)^{3 / 2}-\left(r_{j-1}^{2}-R_{i-1}^{2}\right)^{3 / 2}\right]
.\end{equation}

For the pressure maps, we estimated the noise as the mean pressure in the range [8,10]~Mpc. We subtracted this noise from 2D radial profiles. We then assumed that the background has been appropriately taken into account. The pressure in the outer cylinder times its volume is then equal to the pressure in the external spherical shell times its own volume, so we have

\begin{equation}
    P(R_1,R_2)*\pi(R_2^2-R_1^2)D=P^{\prime}(r_1,r_2)*V_{int}(r_1,r_2,R_1,R_2)
.\end{equation}

With $P(R_1,R_2)$ being the 2D radial pressure in the cylinder extending from $R_1$ to $R_2$ and $P'(r_1,r_2)$ the 3D-deprojected pressure in the volume $V_{int}(r_1,r_2,R_1,R_2)$. $D$ is the diameter of the cluster, it is defined during the random selection of the external limit of the cluster used for the deprojection. We then have the expression for $P^{\prime}(r_1,r_2)$:
 
\begin{equation}
    P^{\prime}(r_1,r_2)=P(R_1,R_2)*\frac{\pi(R_2^2-R_1^2)D}{V_{int}(r_1,r_2,R_1,R_2)}
.\end{equation}Following this reasoning, in the second to last interval we have the equality below, with $V_{int}(r_0,r_1,R_0,R_1)$ being the black area and $V_{int}(r_1,r_2,R_0,R_1)$ the dashed area in the [$R_0$,$R_1$] range in Fig. \ref{dep_scheme}:
\begin{equation}
    P(R_0,R_1)*\pi(R_1^2-R_0^2)D=P^{\prime}(r_0,r_1)*V_{int}(r_0,r_1,R_0,R_1)+P^{\prime}(r_1,r_2)*V_{int}(r_1,r_2,R_0,R_1)
.\end{equation}This leads to 

\begin{equation}
    P^{\prime}(r_0,r_1)=\frac{P(R_0,R_1)*\pi(R_1^2-R_0^2)D-P^{\prime}(r_1,r_2)*V_{int}(r_1,r_2,R_0,R_1)}{V_{int}(r_0,r_1,R_0,R_1)}
.\end{equation}

So we can iteratively go from the outskirt of the cluster to its core, the general equation for any interval is 

\begin{equation}
P^{\prime}\left(r_{i-1}, r_{i}\right)=\frac{P(R_{i-1}, R_{i}) \pi(R_i^2-R_{i-1}^2)D-\sum_{j=i+1}^{m}\left[P^{\prime}(r_{j-1}, r_{j})V_{\mathrm{int}}\left(r_{j-1}, r_{j} ; R_{i-1}, R_{i}\right)\right]}{V_{\mathrm{int}}\left(r_{i-1}, r_{i} ; R_{i-1}, R_{i}\right)}
\label{p_prime_ri}
.\end{equation} 

Given that for the electron density we have column densities (in $m^{-2}$), we need to adapt the method to obtain an electron density (in $m^{-3}$). First, the noise is calculated as the mean of the column density in the range [8,10]~Mpc divided by the depth of the projection $L_{los}$=22,123~Mpc. We have this equation:

\begin{equation}
    n_e^{back}[m^{-3}]=\frac{n_e(8Mpc<R<10Mpc)[m^{-2}]}{L_{los}} 
.\end{equation}

In the outer and the second to last intervals, we have the equations below, respectively, with $n_e^{\prime}$ being the 3D-deprojected electron density in $m^{-3}$ and $n_e$ the electron density in column density in $m^{-2}$:

\begin{equation}
    n_e(R_1,R_2)*\pi(R_2^2-R_1^2)=n_e^{\prime}(r_1,r_2)*V_{int}(r_1,r_2,R_1,R_2)+n_e^{back}*(\pi(R_2^2-R_1^2)*L_{los}-V_{int}(r_1,R_{cluster},R_1,R_2)),
\end{equation}

\begin{equation}
    n_e(R_0,R_1)*\pi(R_1^2-R_0^2)=n_e^{\prime}(r_0,r_1)*V_{int}(r_0,r_1,R_0,R_1)+n_e^{\prime}(r_1,r_2)*V_{int}(r_1,r_2,R_0,R_1)+n_e^{back}*(\pi(R_1^2-R_0^2)*L_{los}-V_{int}(r_0,R_{cluster},R_0,R_1)),
\end{equation}where $R_{cluster}$ is the cluster radius ($R_{cluster}=R_2=r_2$ in this case). Finally, the general equation is 

\begin{equation}
n_{\mathrm{e}}^{\prime}\left(r_{i-1}, r_{i}\right)=\frac{n_{\mathrm{e}}\left(R_{i-1}, R_{i}\right)\pi(R_i^2-R_{i-1}^2)-n_{\mathrm{e}}^{back}(\pi(R_i^2-R_{i-1}^2)L_{los}-V_{int}(r_{i-1},R_{cluster},R_{i-1},R_i))-\sum_{j=i+1}^{m}\left[n_{\mathrm{e}}^{\prime}\left(r_{j-1}, r_{j}\right) V_{\mathrm{int}}\left(r_{j-1}, r_{j} ; R_{i-1}, R_{i}\right)\right]}{V_{\mathrm{int}}\left(r_{i-1}, r_{i} ; R_{i-1}, R_{i}\right)}
.\end{equation}

\newpage

\section{High- and low-resolution pressure projections comparison} \label{appendix_pcompar}

\begin{figure}[h]
   \centering
    \includegraphics[trim=50 200 400 30, width=1\textwidth]{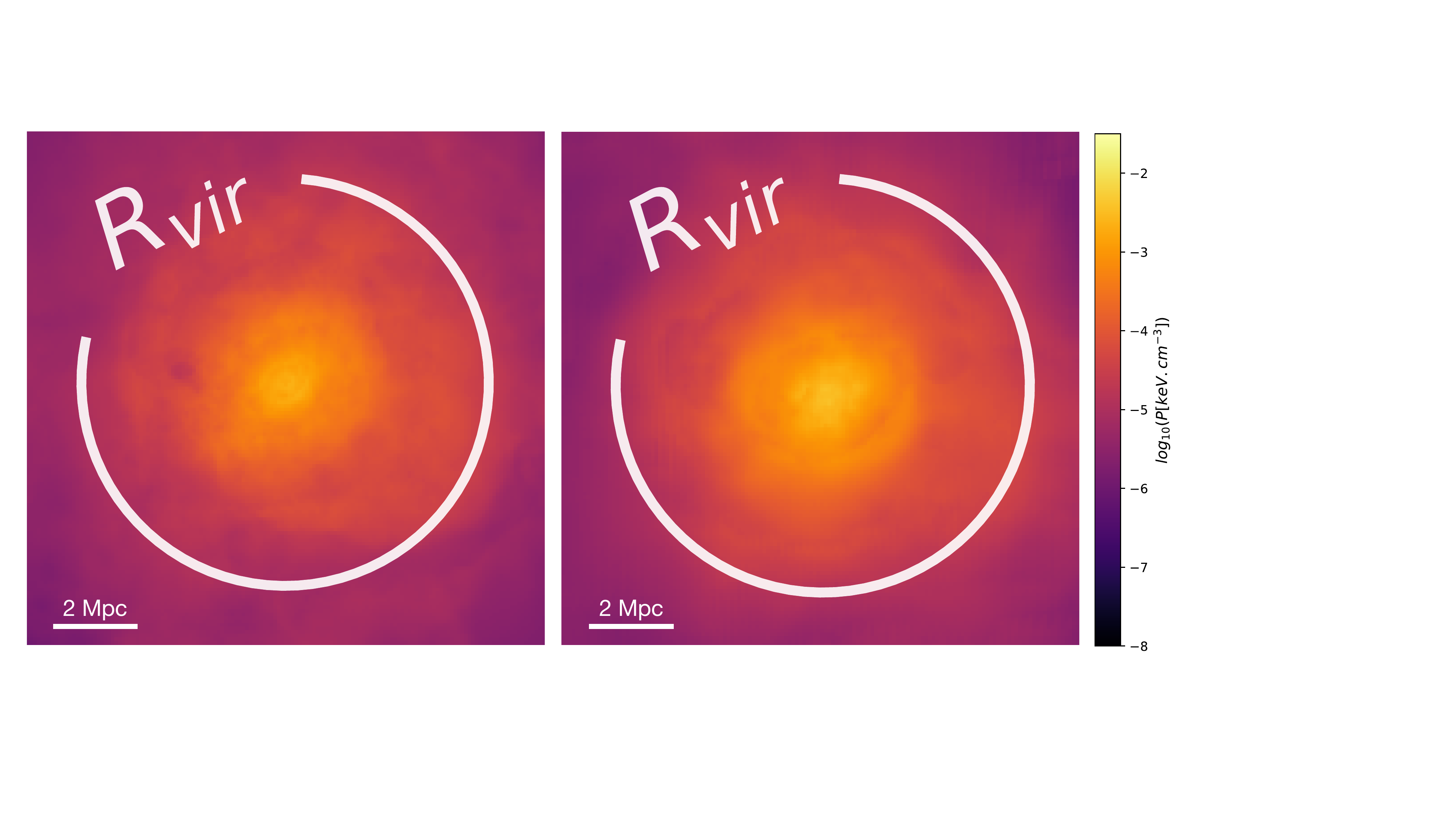}
    \caption{Pressure maps from Fil direction in high- (left) and low-resolution (right) simulations. The maps are $\sim$ 12~Mpc wide and centred on the Virgo cluster centre. The circle is the virial radius.}
      \label{p maps 2 h l}       
\end{figure}

Fig. \ref{p maps 2 h l} presents pressure maps from the Fil direction in the high- (left) and low-resolution (right) simulations. The maps are $\sim$ 12~Mpc wide and centred on the Virgo cluster centre. The circle is the virial radius. It shows that the pressure discontinuity due to the AGN feedback in the cluster's core is more extended in the low-resolution simulation due to numerical diffusion, as discussed in Section \ref{sec:6}.

\end{document}